\documentclass[aps,prl,reprint,superscriptaddress]{revtex4-1}

\usepackage[utf8]{inputenc}
\usepackage{graphicx}
\usepackage{xcolor}
\begin{document}
		\title{Tunable high-harmonic generation by chromatic focusing of few-cycle laser pulses}

	\author{W. Holgado}
	\email{warein@usal.es}
	\altaffiliation{Present Addres: Spanish Center for Pulsed Lasers, M5 Bldg. Science Park, Villamayor, Salamanca, Spain}

	\affiliation{Grupo de Investigaci\'on en Aplicaciones del L\'aser y Fot\'onica, Departamento de F\'isica Aplicada, Universidad de Salamanca, E-37008 Salamanca, Spain}
	
	\author{C. Hern\'andez-Garc\'ia}
	\affiliation{Grupo de Investigaci\'on en Aplicaciones del L\'aser y Fot\'onica, Departamento de F\'isica Aplicada, Universidad de Salamanca, E-37008 Salamanca, Spain}
	
	\author{B. Alonso}
	\affiliation{Grupo de Investigaci\'on en Aplicaciones del L\'aser y Fot\'onica, Departamento de F\'isica Aplicada, Universidad de Salamanca, E-37008 Salamanca, Spain}
	\affiliation{IFIMUP-IN and Departamento de F\'isica e Astronomia, Universidade do Porto, Rua do Campo Alegre 687, 4169-007 Porto, Portugal}
	
	\author{M. Miranda}
	\affiliation{Department of Physics, Lund University, P.O. Box 118, SE-221 00 Lund, Sweden}
	
	\author{F. Silva}
	\affiliation{IFIMUP-IN and Departamento de F\'isica e Astronomia, Universidade do Porto, Rua do Campo Alegre 687, 4169-007 Porto, Portugal}
	\affiliation{Sphere Ultrafast Photonics, Lda, R. Campo Alegre 1021, Edifício FC6, 4169–007 Porto, Portugal}
	
	\author{O. Varela}
	\affiliation{Spanish Center for Pulsed Lasers, M5 Bldg. Science Park, Villamayor, Salamanca, Spain}
	
	\author{J. Hern\'andez-Toro}
	\affiliation{Spanish Center for Pulsed Lasers, M5 Bldg. Science Park, Villamayor, Salamanca, Spain}
	
	\author{L. Plaja}
	\affiliation{Grupo de Investigaci\'on en Aplicaciones del L\'aser y Fot\'onica, Departamento de F\'isica Aplicada, Universidad de Salamanca, E-37008 Salamanca, Spain}
	
	\author{H. Crespo}
	\affiliation{IFIMUP-IN and Departamento de F\'isica e Astronomia, Universidade do Porto, Rua do Campo Alegre 687, 4169-007 Porto, Portugal}
	\affiliation{Grupo de Investigaci\'on en Aplicaciones del L\'aser y Fot\'onica, Departamento de F\'isica Aplicada, Universidad de Salamanca, E-37008 Salamanca, Spain}

	\author{I. J. Sola}
	\email{ijsola@usal.es}
	\affiliation{Grupo de Investigaci\'on en Aplicaciones del L\'aser y Fot\'onica, Departamento de F\'isica Aplicada, Universidad de Salamanca, E-37008 Salamanca, Spain}

	\renewcommand{\bibname}{References}
	\date{\today}
	
	\begin{abstract}
		In this work we study the impact of chromatic focusing of few-cycle laser pulses on high-order harmonic generation (HHG) through analysis of the emitted extreme ultraviolet (XUV) radiation. Chromatic focusing is usually avoided in the few-cycle regime, as the pulse spatio-temporal structure may be highly distorted by the spatiotemporal aberrations. Here, however, we demonstrate it as an additional control parameter to modify the generated XUV radiation. We present experiments where few-cycle pulses are focused by a singlet lens in a Kr gas jet. The chromatic distribution of focal lengths allows us to tune HHG spectra by changing the relative singlet-target distance. Interestingly, we also show that the degree of chromatic aberration needed to this control does not degrade substantially the harmonic conversion efficiency, still allowing for the generation of supercontinua with the chirped-pulse scheme, demonstrated previously  for achromatic focussing. We back up our experiments with theoretical simulations reproducing the experimental HHG results depending on diverse parameters (input pulse spectral phase, pulse duration, focus position) and proving that, under the considered parameters, the attosecond pulse train remains very similar to the achromatic case{, even showing cases of {isolated} attosecond pulse generation for near single-cycle driving pulses.}
	\end{abstract}
	
	\pacs{}

	\maketitle

	\section{Introduction}
	
	Few-cycle pulses are of great interest for attosecond science, allowing the generation of isolated attosecond pulses via high-harmonic generation (HHG) \cite{hentschel2001attosecond,sansone2006isolated}, atomic and molecular dynamic studies \cite{corkum2007attosecond}, ultrafast spectroscopy (e.g., transient absorption \cite{goulielmakis2010real}) and spectral interference in the extreme ultraviolet (XUV) range  \cite{fuchs2012optical}, among others{)}. Nowadays, intense few-cycle pulses in the near-visible to infrared (IR) domain are available thanks to the development of ultrashort pulse lasers combined with post-compression techniques (e.g., based on gas-filled hollow-core fibers (HCF)  \cite{nisoli1996generation,Sartania:97} or filamentation in gases \cite{hauri2004generation}). Proper output spectral phase compensation may directly lead to sub-1.5 cycle pulse compression \cite{silva2014simultaneous}, and special setups  allow to synthesize even sub-cycle pulses \cite{wirth2011synthesized}.
	
	Apart from the complexity of their generation, few-cycle pulses are extremely sensitive to dispersion in the propagation medium and are also prone to spatio-temporal distortions. Thus, in order to preserve the duration and spatio-temporal properties of few-cycle pulses, focusing with achromatic and non-dispersive systems, such as spherical mirrors or off-axis parabolic reflectors \cite{alonso2013spatiotemporal}, is required. However, few-cycle pulses obtained by post-compression techniques typically exhibit spatio-temporal structure. In the case of filamentation this is more evident \cite{alonso2011spatiotemporal}, but spatial dependence is also present when using the gas-filled HCF technique, namely in the form of spatial chirp \cite{alonso2013characterization}. Incidentally, controlling the spatio-temporal structure of the beam may be used to exert additional control over nonlinear light-matter interaction processes. Using diffractive optical elements (DOEs) \cite{alonso2012frequency} or lenses exhibiting some aberrations, is a simple way to introduce a spatio-temporal structure in ultrashort laser pulses. For instance, chromatic focal aberrations {allows tuning} the second harmonic wavelength by simply adjusting the distance between the nonlinear crystal and the focusing element \cite{minguez2010wavelength}. Also, chromatic focusing can be used to tune the broad spectrum resulting {from} a filamentation process \cite{borrego2013femtosecond}. In addition, including astigmatic focusing allows for the generation of more stable, spectrally broader, higher energy filaments \cite{alonso2011enhancement} than for the non-astigmatic case. Aberrated focusing is also used to improve the axial resolution and to extend the penetration depth in nonlinear confocal microscopy \cite{gualda2010wavefront}.

	The control of HHG through the introduction of aberrations to the fundamental beam would have practical implications in  fields such as XUV spectroscopy or the temporal shaping of attosecond pulses. In fact, pulse front tilt has already been proven as an useful tool to generate angle-dependent XUV radiation emission \cite{wheeler2012attosecond}, allowing one to spatially filter isolated attosecond bursts. Similarly, angular chirping of the driving field can be used to generate XUV radiation with controlled angular distribution of the spectra \cite{hernandez2016high}. Therefore, what is an \textit{a priori} detrimental aberration can be turned into a useful tool. Furthermore, several efforts have been devoted to generate tunable radiation in the XUV range, for instance by tuning the driving pulse frequency \cite{shan2002tunable,reitze2004enhancement} or altering the gas parameters in the generation \cite{lu2013generation}. Brandi \textit{et al} show that the tuning of XUV radiation can lead to a maximum absorption in certain atomic transitions \cite{brandi2003high}.

	In this work we explore experimentally and theoretically high-order harmonic generation driven by few-cycle pulses chromatically focused with a normal-dispersion convergent singlet lens. Chromatic aberration affects the driving IR light distribution at the focal region {since} focal length is shorter (longer) for shorter (longer) wavelengths. As a result, we introduce a new degree of control of the XUV spectrum characteristics. In particular, we demonstrate that it is possible to shift the XUV spectrum and to modify the spectral content while maintaining the yield and attosecond train structure similar to the obtained with achromatic focusing. The experiments are complemented and corroborated by numerical simulations of the focusing scheme in a macroscopic gas sample. This paper is organized as follows: firstly, the experimental and theoretical procedures are described; secondly, the experimental results are shown; then, the theoretical results are presented, discussing the role of the chromatic focusing as a new control parameter of the HHG process.

	\section{Experimental and theoretical methods}

	For the experiments we used a 1-kHz Ti:Sapphire CPA amplifier (Femtolasers FemtoPower Compact Pro CEP) delivering pulses with a Fourier-transform limit duration of 25 fs of full-width at half-maximum (FWHM). The output pulse is post-compressed in a HCF with an inner diameter of 250 micrometers and 1-meter length. The HCF was filled with {a}rgon at 1 bar. By compensating the spectral phase with 10 bounces off chirped mirrors (Ultrafast Innovations; nominal GDD: $-20\ \mathrm{fs}^2$ per bounce at 800 nm, minimum reflectance: 99\%), 5-fs (and shorter) pulses with an energy up to 300 $\mu$J are routinely obtained \cite{silva2014simultaneous,alonso2013characterization}. Pulse duration can be tuned by changing the post-compression gas pressure and re-adjusting the spectral phase compensation. In the present work, pulse duration covered the range between 3.3 and 8.0 fs.  
	
	The laser pulse was then focused into a krypton gas jet by a BK7 glass singlet lens ($f = 30$ cm at $\lambda$ = 800 nm), which provided the desired chromatic focusing scheme. We also performed experiments with an achromatic focusing scheme for comparison. For the latter we employed a spherical silver mirror ($f = 50$ cm). Both the lens and the spherical mirror were placed on a translation stage, so the focus position could be controlled and scanned. Before the focusing system, a variable aperture is used to optimize the HHG signal, where the diameter of the beam is truncated to 5 mm, and optimized before each scan. The pulse entered the vacuum chamber through a 0.5 mm thick fused-silica window, which was placed close to the focusing element to avoid any potential nonlinear effects. HHG was performed in a krypton gas jet (5 bar of backing pre\-ssure), with a nozzle of 500 $\mu$m diameter. The pressure inside the vacuum chamber where the high-order harmonics were generated was around 5$\times10^{-3}$ mbar. A 150-nm thick aluminum foil was used to filter out the IR radiation and the lower-order harmonics, while the higher orders (with energies between 17 eV and 70 eV) {were propagated} through it.
	
	The XUV spectra were characterized with a grazing-incidence Rowland circle XUV spectrometer (Model 248/310G, McPherson Inc.), of 1-m radius, equipped with a 300 grooves/mm spherical diffraction grating. {The {maximum} detection angle in the present configuration is 2 mrad, while the XUV radiation divergence is estimated to be 1 mrad, with the final spectrum being the integration over all angles of propagation}. The carrier-envelope phase (CEP) of the seed oscillator (Femtolasers Produktions Rainbow CEP) was stabilized with a fast loop and its stability was not significantly altered by the subsequent amplification and post-compression processes, which resulted in an rms of approximately 100 mrad throughout each measurement without the need to employ a slow loop \cite{holgado2016continuous}. The driving few-cycle laser pulses were temporally characterized using the d-scan technique \cite{silva2014simultaneous,miranda2012characterization}, which can measure pulses down to single-cycle durations \cite{silva2014simultaneous,Miranda:17}.

	\begin{figure*}
		\includegraphics[width=1\linewidth]{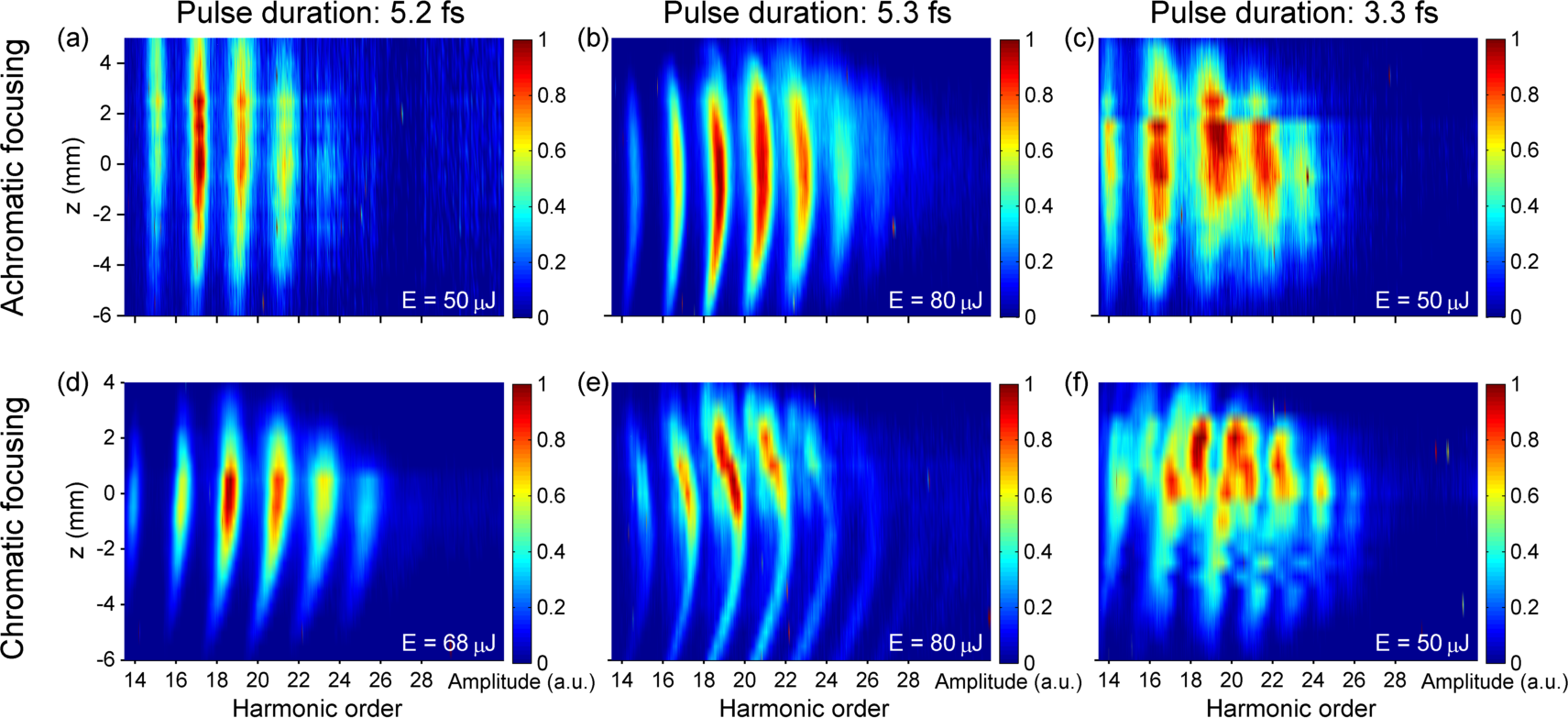}
		\caption{\label{zscans} \textbf{HHG $z$-scans for chromatic and achromatic case:} HHG {in} Kr depending on focusing system-target distance $z$ ($z = 0$ mm stands for focus on gas jet, $z < 0$ for focus before gas jet, $z>0$ for focus after gas jet) for three different driving pulses (shown in Fig. 2): (a, d) 5.2 fs FWHM pulse (Fourier limit: 3.35 fs) with residual third order dispersion, (b,e) 5.3 fs FWHM pulse (Fourier limit: 4.9 fs) and (c,f) 3.3 fs FWHM pulse (Fourier limit: 2.9 fs). First row figures (a-c) correspond to focusing with a silver spherical mirror ($f = 50$ cm), and the lower row (d-f) to focusing with a BK7 singlet ($f = 30$ cm).}
	\end{figure*}

	We performed numerical simulations of HHG including both microscopic (single-atom) and macroscopic (phase-matching) calculation. The dipole acceleration of each elementary emitter was calculated using an extension of the strong field approximation \cite{perez2009s} (due to the lack of simple analytical expressions for the matrix elements of the acceleration in krypton, we have performed our calculations in argon, that has the same parity for the valence electron and a similar ionization potential), while harmonic propagation and the collective response were worked out using a method based on the electromagnetic field propagator \cite{hernandez2010high}. The gas jet was modelled with a Gaussian density distribution along the laser propagation direction with a width of 1-mm (FWHM) and a peak pressure of 10 mbar. {For the parameters used in our calculations, the ionized population estimated from the Ammosov-Delone-Krainov theory \cite{adk1986} {is} approximately 10\%.}
	
	The signal at the far field detector was computed as the coherent addition of the HHG contributions of all the elementary sources, where the HHG light was assumed to propagate to the detector with a phase velocity $c$ (vacuum velocity of light). In this process, phenomena such as time-dependent group velocity walk-off  \cite{hernandez2016group}, absorption of the harmonics, plasma and neutral dispersion were all taken into account. At the pressures used in this work, nonlinear phenomena in the propagation of the driving IR beam were negligible. This model has been tested on several scenarios where phase matching in a gas jet is a relevant factor in HHG \cite{holgado2016continuous,hernandez2015carrier,hernandez2013signature,kretschmar2013spatial,hickstein2015non,hernandez2013attosecond}.
	
	In this work, and in order to simulate experiments where the driving beam was focused with a singlet lens, the chromatic effect was  also taken into account. The refractive index $n(\lambda)$ of BK7 is obtained from Sellmeier equations, which is used to finally obtain the  {wavelength-dependent focal length,} $f(\lambda)={R}\cdot{(n(\lambda)-1)}^{-1}$, in the whole spectral bandwidth of the fundamental pulse, where $R$ is {the} radius of curvature of the singlet. Once the focal length dependence $f(\lambda)$ has been obtained, the propagation after the singlet at different distances around the central focal length was calculated. For this purpose, we propagated each wavelength using Fresnel's diffraction as described in Ref. \cite{alonso2012frequency}.

	\section{Experimental results}
	When focusing broadband pulses with a system exhibiting chromatic aberration, the focal length depends on the wavelength. Thus, if such a beam is used for driving a nonlinear process,  {such} as HHG, the spectrum of the generated radiation will strongly depend on the relative position of the medium with respect to the nominal beam focus. In this situation, the XUV spectra will be modified by the lens position {along} coordinate $z$. 
	
\begin{figure}[b]
			\includegraphics[width=1\linewidth]{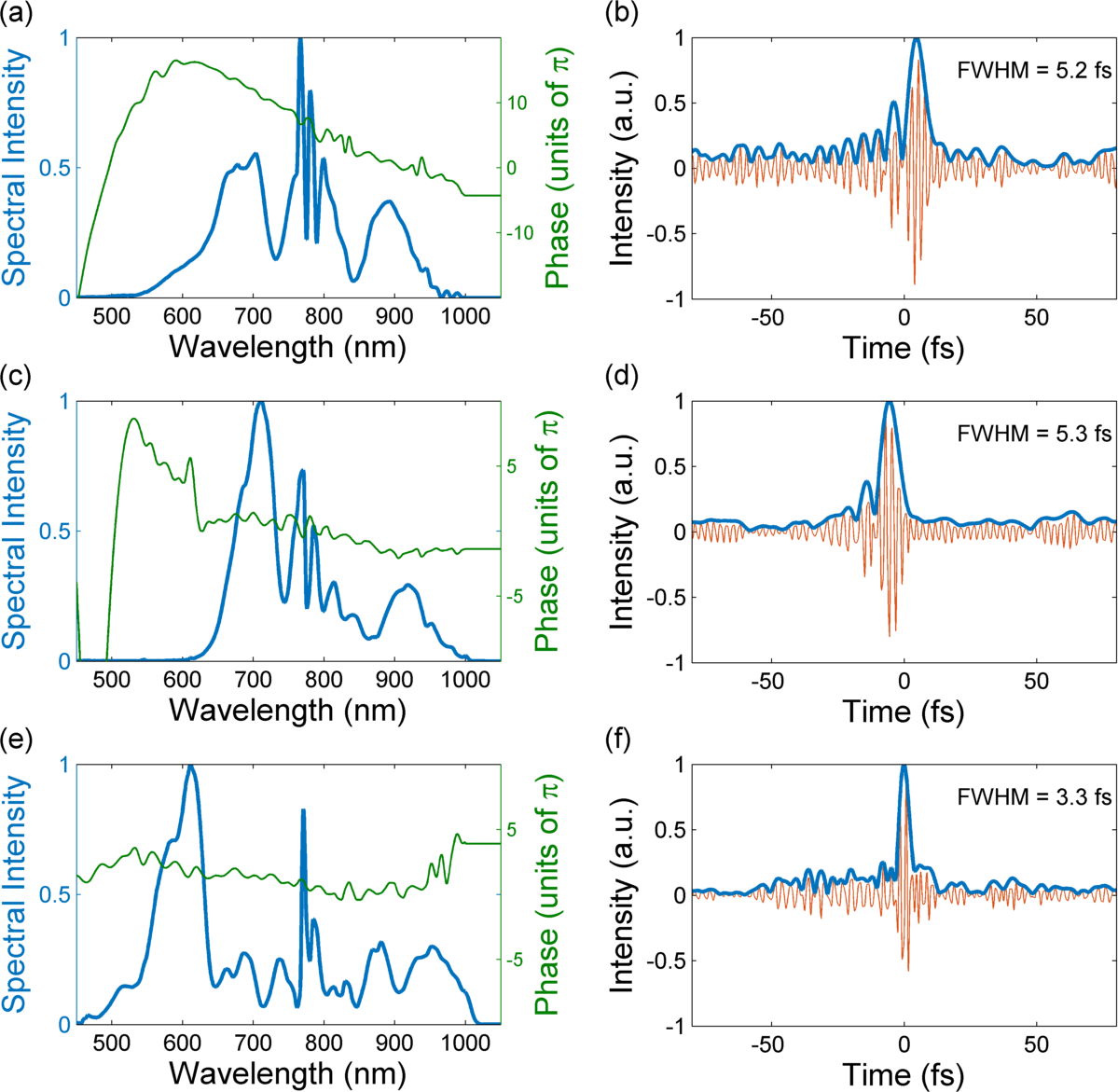}
			\caption{\label{dscans} \textbf{Characterization of the driving pulses:} Left column: Spectra (blue line) and spectral phase (green line); right column: time dependent electric field amplitude (blue line) of the three pulses used for the HHG $z$-scan shown in Fig.\ \ref{zscans}. Upper row shows a pulse with higher dispersion order, and corresponds to Fig.\ \ref{zscans}(a) and (d). Second row displays a pulse closer to its Fourier-limit duration, where the spectral phase is flatter than in the previous case, and corresponds to Fig. \ \ref{zscans}(b) and (e). Lower row presents a shorter pulse, lasting less than 1.5 cycles, which corresponds to Fig.\ \ref{zscans}(c) and (f).}
		\end{figure}

	To confirm the latter point, we measured the harmonic spectra varying the distance ($z$ coordinate) between the focusing element and the gas jet (we shall refer to this as HHG $z$-scan, not to be confused with the nonlinear material characterization technique called $z$-scan). The $z=0$ position here stands for focus on gas jet, $z<0$ for focus before gas jet and $z>0$ for focus after gas jet. Fig{ure}\ \ref{zscans} shows a set of HHG $z$-scans {over} different input pulse conditions (5.2 fs FWHM pulses, for figures \ref{zscans}a and \ref{zscans}d; 5.3 fs FWHM pulses, with {optimized spectral phase} for figures \ref{zscans}b and \ref{zscans}e and 3.3 fs FWHM pulses for figures \ref{zscans}c and \ref{zscans}d) and focusing schemes (upper row corresponds to focusing with the spherical mirror and lower row {to} singlet lens focusing). The input pulse {d-scan} reconstructions corresponding to the considered cases are shown in Fig.\ \ref{dscans}.
	
	Fig{ure}\ \ref{zscans}a confirms that, when a spherical mirror is used, the spectral position of the harmonics remains unaltered when the distance between the focusing mirror and the gas jet is change{d}. This {is to be expected,} since no chromatic aberration is present.  In this case, the 5.3 fs FWHM input pulse exhibits residual TOD (see Fig{s}.\ \ref{dscans}a and \ref{dscans}b), {which} is very common after {the} post-compression process and is caused by the nonlinear process inside the hollow-core fiber under optimized propagation conditions \cite{conerejo2016tod}, {as} previously observed in several works \cite{silva2014simultaneous,alonso2011enhancement,fabris2015single,heyl2016scale,bohle2014compression}.

	In Fig.\ 1b the same scan was performed with a different pulse. In this case the spectral broadening during the HCF post-compression was reduced, the pulse duration was nevertheless maintained at 5.3 fs FWHM (Figs.\ \ref{dscans}c and \ref{dscans}d) due to optimization of the higher orders of the pulse spectral phase. In these conditions, we see that the HHG $z$-scan varies from the previous $z$-coordinate independent case, since some spectral broadening, in particular at higher harmonics, is observed at $z>0$ focusing positions (Fig.\ \ref{zscans}b). This spectral broadening, as recently reported in Ref. \cite{holgado2016continuous} at the same pulse duration regime, results from a joint effect at the atomic (HHG is altered by particular input pulse spectral phases) and macroscopic levels (due to collective coherent addition, phase matching fills the harmonic spectrum, enhancing the continuum structure and smoothing out spectral peaks).

	Simultaneously, a blue-shift appears in the harmonics. We explain this effect as the result of a remaining chromatic effect that {occurs} when a beam is focused with an achromatic optical element (i.e., a mirror), as explained in Ref. \cite{Karimi:13}. These differences between the two scans demonstrate the effect of the spectral phase of the driving pulse in the HHG process.

	Furthermore, Fig.\ \ref{zscans}c shows that the spectral broadening at $z>0$ focusing position is increased when using shorter pulses (3.3 fs FWHM, whose reconstruction{s} are shown in Fig{s}.\ \ref{dscans}e and \ref{dscans}f). {In this last case, even a continuum spectrum can be obtained due to the short driving pulse duration.}

		\begin{figure}[t]
			\includegraphics[width=1\linewidth]{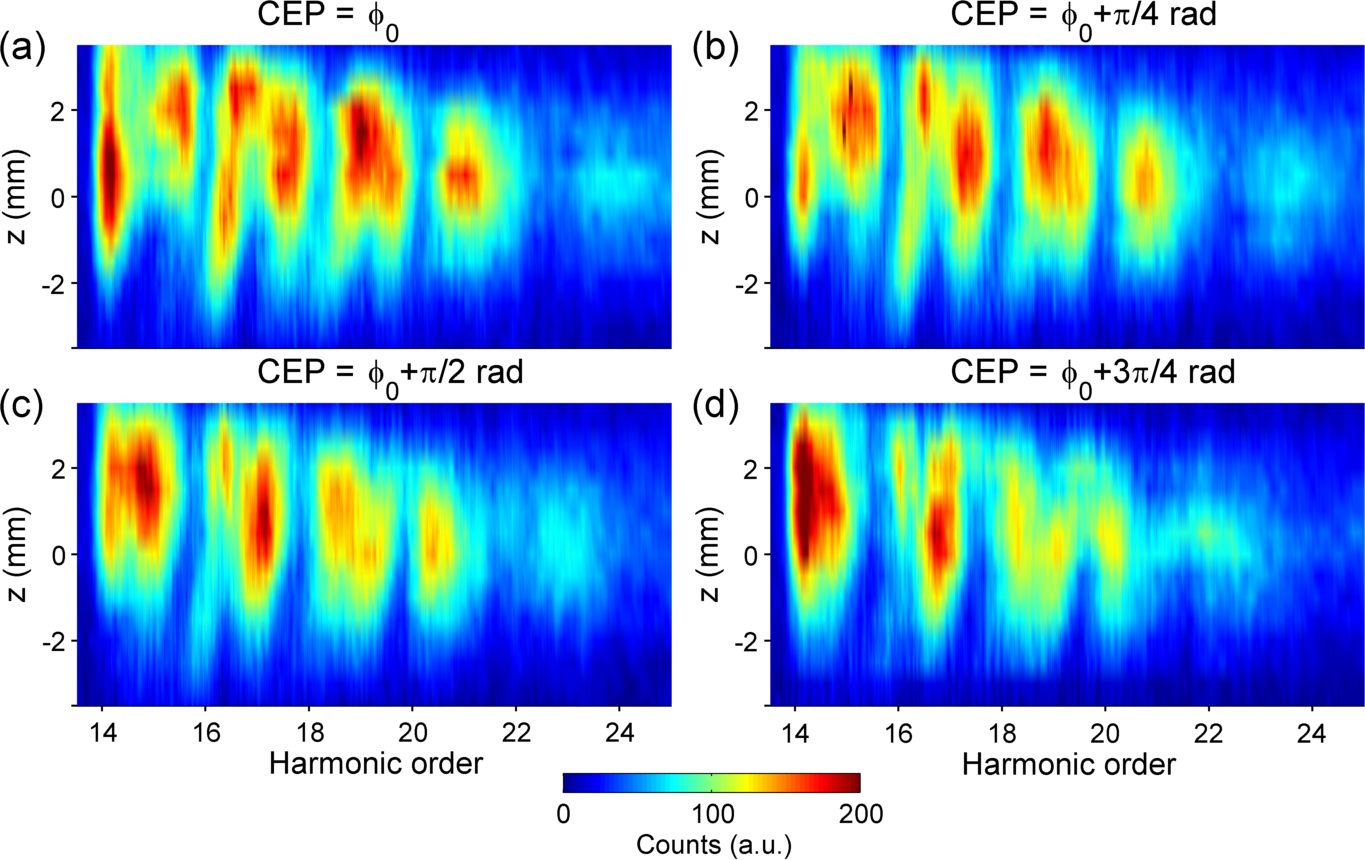}
			\caption{\label{zscanscep} \textbf{CEP-dependence of the chromatic HHG $z$-scan}: HHG $z$-scan {in} Kr (backing pressure of 6 bar) as a function of singlet-target distance in the few-cycle pulse regime (pulse energy of 56 $\mu$J, {pulse} duration: 3.6 fs FWHM, Fourier limit: 2.9 fs FWHM) for different values of CEP: (a) $\phi_0$, (b) $\phi_0+\pi/4$, (c) $\phi_0+\pi/2$, (d) $\phi_0+3\pi/4$.}
		\end{figure}

	In order to analyze the chromatic focusing effect, Fig.\ \ref{zscans} lower row presents the corresponding HHG $z$-scan measured while focusing with the singlet. Fig.\ \ref{zscans}d reveals that the chromatic focusing scheme introduces a dependence of the harmonic position on $z$, in contrast with was observed using achromatic focusing (Fig.\ \ref{zscans}a) and the same input pulse characteristics. In fact, it follows the first intuitive guess: when moving the lens in order to place the nominal focus after (before) the gas jet, i.e. $z>0$ ($z<0$), a blue shift (red shift) of the harmonic spectrum should be observed. Since no additional effect{s} alter the XUV spectra, this dependence is observed clearly. However, the presence of the above commented spectral broadening phenomenon will alter this behavior under other experimental conditions. As shown in Fig.\ \ref{zscans}e, for $z>0$, XUV spectra become broader and {their} dependence {in} $z$ is opposite {to what would be} expected from {a purely} chromatic effect. Nevertheless, when the {spectral} broadening effect is not present, i.e. at $z<0$, the spectral shift induced by the chromatic focusing prevails, shifting the {measured} harmonic spectral positions away from the corresponding achromatic focusing case (Fig.\ \ref{zscans}b). This occurs similarly when shorter input pulses {of} 3.3 fs FWHM are used, since at $z>0$ spectral broadening is present, {and} even enhanced (e.g., continuous spectra arise), but for $z<0$ the spectral shift matches {that of} the used chromatic focusing, in contrast also with {the insensivity in} the $z$ {coordinate} observed in the achromatic case (Fig.\ \ref{zscans}c). Thus, chromatic focusing effectively allows one to vary the HHG spectrum and to tune it by just moving the lens position {with respect to the gas jet}.

	\begin{figure*}
		\includegraphics[width=1\linewidth]{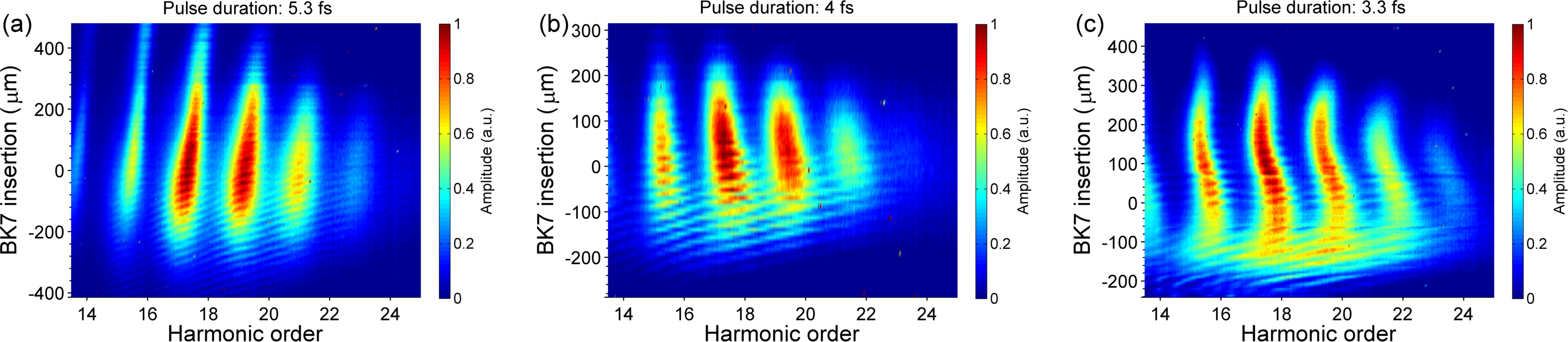}
		\caption{\label{chirpscans} \textbf{HHG dispersion-scans with chromatic focusing:} HHG performed in krypton in three different cases: (a) Pulse duration of 5.3 fs and energy of 80 $\mu \mathrm{J}$, (b) duration of 4 fs and 50 $\mu \mathrm{J}$, (c) duration of 3.3 fs and 50 $\mu \mathrm{J}$. Focus is placed 2 mm after the gas jet in (a) and (b), while it is placed at the gas jet position in (c). Pulse durations are measured at 0 $\mu$m of BK7 insertion. Color axis is normalized to the maximum signal measured in each scan}
	\end{figure*}

	The HHG $z$-scan in the few-cycle pulse regime is extremely sensitive to experimental conditions. As an example, Fig. \ref{zscanscep} shows the harmonic {z-scan} intensity profile for several CEPs of the driving pulse (pulse energy of 56 $\mu$J, pulse duration: 3.6 fs FWHM, Fourier-limit: 2.9 fs FWHM). In this regime, due to the chromatic focusing, pulse structure (even at the level of the electric field oscillation) has great relevance, and phase changes in the electric field of the pulses directly affect the HHG $z$-scan.

	In order to study the effects of chromatic focusing in the XUV continuum generation, we have performed a dispersion-scan for the different studied cases. The coordinate $z$ (i.e.{, the} position of the lens) has been chosen to optimize the spectral broadening. A dispersion-scan was performed by adding or extracting material on the light path prior to the focusing element. In our case, we used a pair of BK7 {glass} wedges with an angle of 8º. In the case of the 5.3 fs FWHM pulse (Fig. \ref{dscans}d) used for Fig. \ref{zscans}e, the $z$ coordinate was fixed at $z = 1$ mm and a dispersion-scan was performed (Fig. \ref{chirpscans}a). The scan shows a behavior very similar to that observed when focusing the same pulse achromatically (with the spherical mirror of $f = 50$ cm), as shown in Fig. 2b in \cite{holgado2016continuous}. For positive chirp the HHG spectra feature narrow peaks at odds harmonics, while for negative chirp, spectral broadening is observed. The observed high frequency fringes denote the CEP dependence of the HHG (i.e., {over} a short range of BK7 {insertion} the dispersion-scan becomes a CEP scan). This behavior of the XUV radiation has been observed pre\-vious\-ly in other works using achromatic focusing \cite{wang2008generation,rudawski2015carrier}.

	When shorter pulses (e.g., 4 fs) are used (Fig.\ \ref{chirpscans}b){, dispersion-dependent spectral} changes are more pronounced, from well-defined harmonics at positive chirp to spectral broadening at negative chirp, until eventually becoming continua, ranging over 8 harmonic orders, {similar to the previously} reported case for achromatic focusing \cite{holgado2016continuous}. This suggests that the continuum mechanism remains the same as for achromatic focusing {in} this pulse duration range. However, shorter pulses in the few-cycle regime reveal differences in the dispersion-scan. Fig. \ref{chirpscans}c features a dispersion-scan for a 3.3 fs pulse (Fig. \ref{dscans}f), where the $z$ coordinate was fixed at $z = 0$  mm (the corresponding $z$-scan is depicted in Fig. \ref{zscans}f). In this case, the continuum spans over 12 harmonic orders and, in contrast with the precedent examples, the harmonic {continuum} yield is higher than the one obtained at the same scan in other chirp conditions.  {As {will} be discussed in the following section, when generating high order harmonics with near single-cycle driving pulses, the continuous structure is compatible with the generation of a single attosecond burst.}

	\section{Theoretical results and discussion}
	
	{Aiming to understand the role of the singlet lens in more detail, we studied the effect of chromatic focusing on the HHG, i.e., how chromatic aberration of the driving beam affects the harmonic emission at the microscopic (single-atom) and macroscopic ({propagation and} phase-matching) levels. Numerical simulations have been performed by scanning the pulse duration and lens-gas distance.}

	\begin{figure}[b]
		\includegraphics[width=1\linewidth]{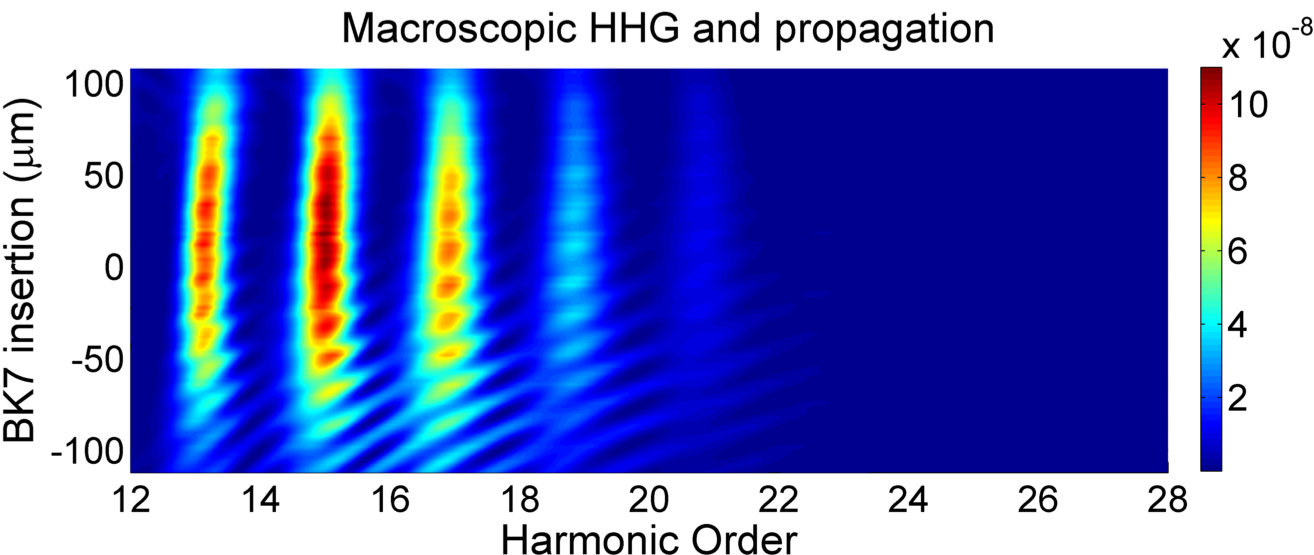}
		\caption{\label{chirpscans_th} \textbf{Numerical simulation of HHG with chromatic focusing:} Dispersion-scan of far field on-axis HHG for Fourier limited pulse of 4.8 fs FWHM focused with a $f = 40$ cm singlet lens. Single atom response and collective response including propagation effects are considered.}
	\end{figure}

	Fig.\ \ref{chirpscans_th} shows the HHG spectra corresponding to the dispersion-scan of a 4.8 fs pulse {(Fourier limit for a BK7 insertion of 0 $\mu$m)}, focused by a $f = 40$ cm singlet and with the nominal focus placed 2 mm after the gas jet, (i.e., $z=2$ mm) {for both} single atom response and collective response. Our simulations have a reasonably good qualitative agreement with experimental results at similar conditions (Figs.\ \ref{chirpscans}a and \ref{chirpscans}b).
	
	\begin{figure}[t]
			\includegraphics[width=1\linewidth]{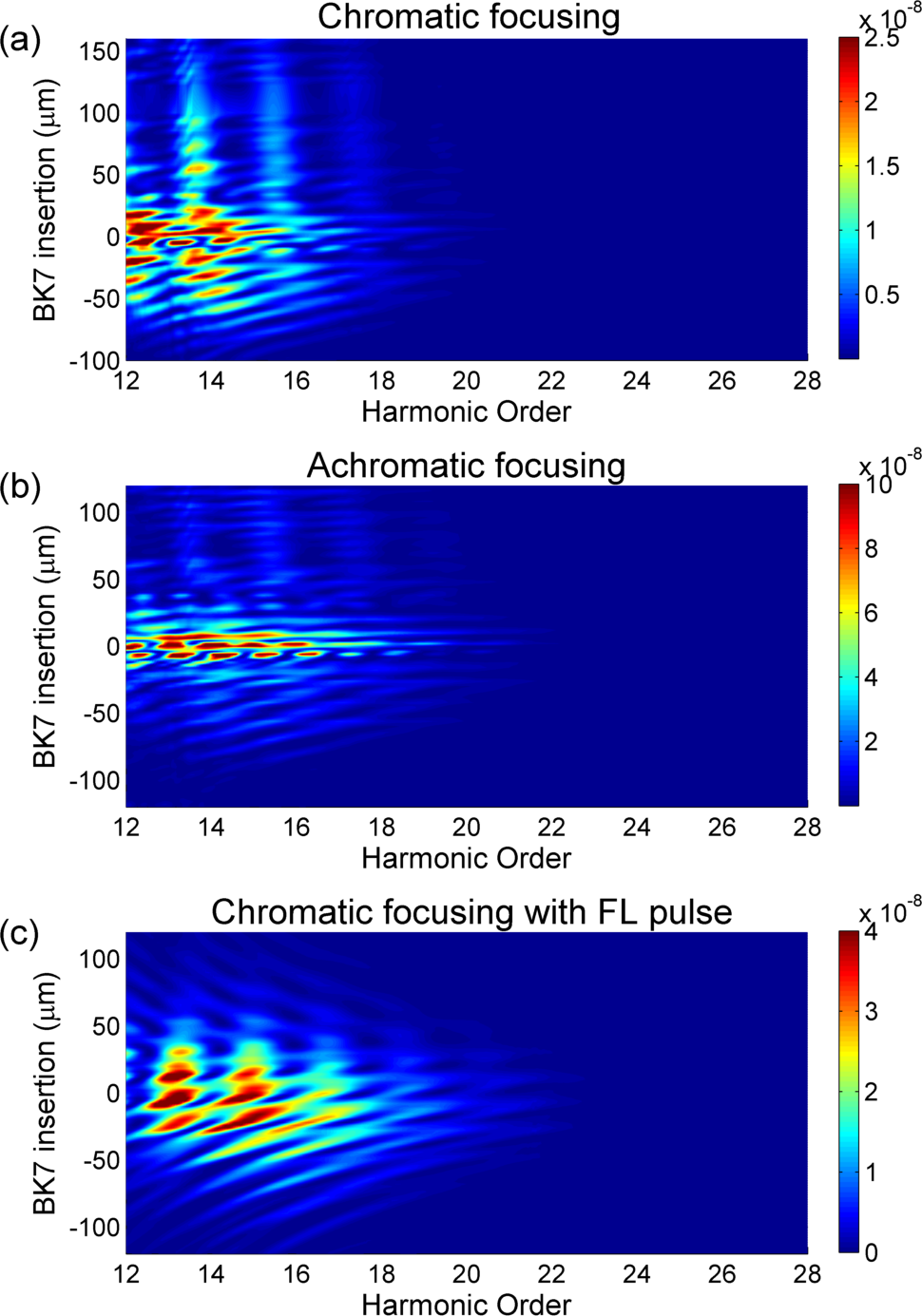}
			\caption{\label{chirpscans_shorterth} \textbf{HHG dispersion-scan impact of focusing scheme and pulse structure:} { Dispersion-scan of far field on axis HHG for Fig. 2f pulse focusing with a $f = 40$ cm singlet at 2 mm before the gas jet (a), same pulse focused with a spherical mirror ($f = 40$ cm) at 2 mm after the gas jet (b) and a 2.9 fs FWHM Fourier Limit pulse focusing with a $f = 40$ cm singlet at 2 mm before the gas jet (c). Maximum peak intensities are estimated to be 2.5$\times 10^{14}\ \mathrm{W}/\mathrm{cm}^2$ for chromatic focusing (a), and 3.5$\times 10^{14}\ \mathrm{W}/\mathrm{cm}^2$ for the achromatic case (b)}.}			
	\end{figure}	
	
	When comparing the chromatic focusing case {shown} here {with the} achromatic one {for} the same conditions {presented in} \cite{holgado2016continuous}, {we see that} while the dispersion-scans for single atom response present some differences between both focusing schemes, {due to} the different local pulse distribution, the corresponding macroscopic response{s} (Fig.\ \ref{chirpscans_th} versus Fig.\ 3b of \cite{holgado2016continuous}) are very similar in their general structure and yield. Surprisingly, phase matching when focusing with the singlet and its associated chromatic aberration can perform slightly better than aberration-less focusing. According to simulations, the temporal structure of the XUV continuum is the same than the one obtained for spherical mirror focusing  \cite{holgado2016continuous}: three main attosecond pulses with relative phase differences that yield, via spectral interference, the continuum spectrum. Therefore, no major attosecond pulse time structure distortions would be produced by the considered chromatic aberration compared to the achromatic focusing case {in} this driving pulse regime.

	For gaining further insight on the HHG response to the dispersion-scans in the few-cycle pulse regime and its dependence on spectral phase and focusing conditions, we also performed simulations with the shorter pulses (and broader spectra). We simulated the HHG dispersion scans (Fig.\ \ref{chirpscans_shorterth}a) considering the 3.3 fs (FWHM) experimentally reconstructed pulses (Fig. \ref{dscans}e and \ref{dscans}f) when focusing with the singlet with its nominal focus 2 mm after the gas jet ($z = 2$ mm). The HHG dispersion scan structure presents a similar behavior to the experimental results measured for the same input pulses (Fig. \ref{chirpscans}c). A broader spectrum structure appears around the zero GDD and negative GDD region, with continuous spectra arising at several material insertion values. On the other hand, odd harmonic peaks rise for positive BK7 insertion, being quite resilient to dispersion within more than 100 $\mu$m material insertion range. Please note that the odd harmonic peaks show a slight blue shift from the expected values, because we are considering a bluish focus at  $z=2$ mm position. The  shift in {the vertical} axis (BK7 insertion) of the HHG dispersion scan in the experimental measurements {with respect to} the theoretical results may be caused by a slight mismatch {in the amount of} dispersive material {between} the HHG experiment {and} the d-scan measurement.

   In order to establish the role of the focusing scheme and the input pulse {shape} over the structure of the HHG dispersion scan at 3.3 fs, we have performed the same simulations {when} focusing the experimentally retrieved pulses with a mirror ($f=40$ cm), i.e.{, using an} achromatic scheme (Fig. \ref{chirpscans_shorterth}b) and considering singlet focusing of a 2.9 fs FL pulse at 0 $\mu$m BK7 material insertion (Fig. \ref{chirpscans_shorterth}c). In the case of achromatic focusing (Fig. \ref{chirpscans_shorterth}b), even if a structure similar to Fig. \ref{chirpscans_shorterth}a is observed, the material insertion range {for efficient HHG} is limited {to} around {zero insertion}, dropping quickly when {more} BK7 is added or subtracted from that value. This fact points out that the chromatic focusing affects the structure of the input pulse, as expected. In the achromatic focusing case, since the spectrum at the considered focal plane is broader than in the chromatic case, the pulse will exhibit a higher sensitivity {to} the material insertion (i.e., dispersion). Thus, efficient HHG generation is restricted to the low dispersion cases within the chirp-scan. In addition, since that chromatic aberration will decrease the peak power on focus, {the} HHG yield in the zero insertion region is higher in the achromatic case than {in} the singlet one.

	When considering a 2.9 fs FWHM FL driving pulse focused with the singlet (Fig. \ref{chirpscans_shorterth}c) at the same focus position ($z = 2$ mm), the yield of the XUV radiation is comparable to Fig. \ref{chirpscans_shorterth}a, while it is lower than the achromatic focusing case (Fig. \ref{chirpscans_shorterth}b), because of the decreased peak power due to the chromatic aberration. Harmonic generation is restricted to a material insertion range between -50 $\mu m$ and 50 $\mu m$, {which is} broader than for achromatic focusing (Fig. \ref{chirpscans_shorterth}b) but not exhibiting the structure at higher positive insertion shown by the experimental case (Fig. \ref{chirpscans}c) and its corresponding simulation (Fig. \ref{chirpscans_shorterth}a). Thus, the  asymmetric structure observed in the latter cases is due to the spectral phase of the driving pulse.

	{Similarly to what happens with second harmonic generation (SHG) d-scan, nonlinear signal generation is more confined within a certain material insertion range in a FL input pulse than in an input pulse exhibiting third (or {higher}) order dispersion. Due to the structured spectral phase of the pulse, its {dependence on} the applied dispersion will differ from {one} wavelength to another, stretching the d-scan trace {along} the material insertion axis). Secondly, the structure around the zero dispersion value becomes more regular in the FL input pulse case than {that} shown {in} Fig. \ref{chirpscans_shorterth}a for the measured pulse with structured spectral phase, and {is} symmetric {with respect to} the material insertion. This difference is explained also by the different spectral phases and their behavior while changing the material insertion. In addition, Fig. \ref{chirpscans_shorterth}b shows that in an achromatic focusing case, the {structure in} Fig. \ref{chirpscans_shorterth} would be washed out, pointing out that the chromatic focusing plays a role maintaining that scan structure. Thus, from comparison among the 3 cases, we can conclude that the particular scan structure observed experimentally (Fig. \ref{chirpscans}c) and theoretically (Fig. \ref{chirpscans_shorterth}a) is caused by both {the} driving pulse spectral phase structure and {the} focusing conditions. Nevertheless, in spite of the differences, continuous spectra can be observed at some points of all the simulated HHG dispersion scans.}

	\begin{figure}[b]
		\includegraphics[width=1\linewidth]{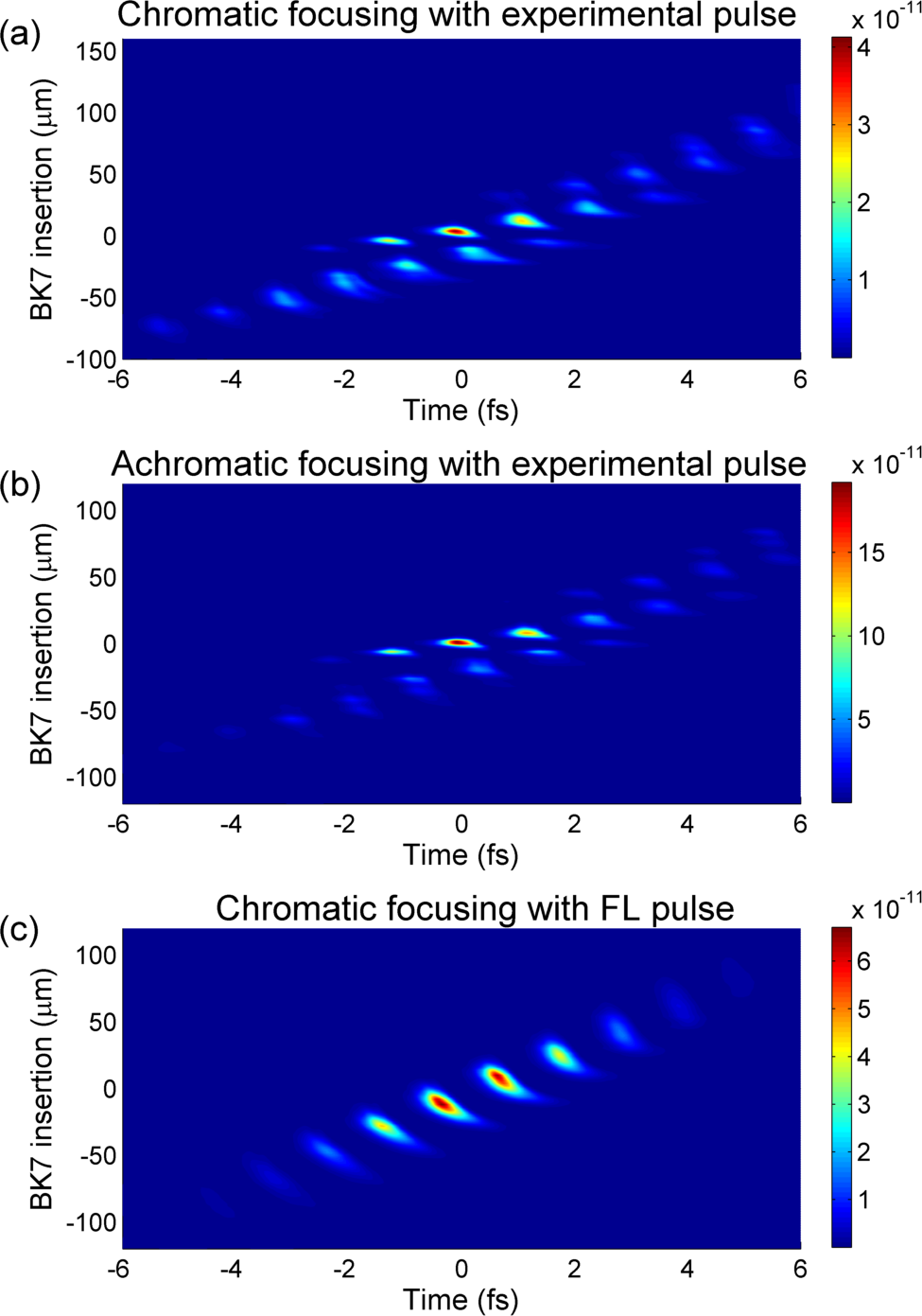}
		\caption{\label{attochirpscans_shorterth} \textbf{ Attosecond pulse emission along the HHG dispersion scan:} {  Attosecond pulse emission corresponding to Fig.\ \ref{chirpscans_shorterth}a (a) Fig.\ \ref{chirpscans_shorterth}b (b) and Fig.\ \ref{chirpscans_shorterth}c (c) HHG dispersion scans. Temporal profile in each case  is obtained by means of Fourier transforming the simulated spectrum after simulating the propagation through the aluminum filter.}}		
	\end{figure}

	{In addition to the spectral analysis, theoretical simulations also allow {studying the temporal structure of the emitted XUV} radiation {when} using near single-cycle pulses (3.3 fs). Contrary to what is observed in the 4.8 fs FWHM driving pulse case, where continuous spectra were generated by a train of 3 attosecond pulse{s} presenting particular spectral phase differences, in this shorter driving pulse case (near single-cycle pulses) some of the continuous spectra correspond to isolated attosecond pulses.  In Fig.\ \ref{attochirpscans_shorterth}a we present the attosecond pulses obtained after Fourier Transforming the spectra shown in Fig.\ \ref{chirpscans_shorterth}a, proving that the measured HHG spectra (Fig. \ref{chirpscans}c) at some material insertion values are compatible with the generation of an isolated attosecond pulse. This does not differ too much from the attosecond pulse emission corresponding to the achromatic focusing case (Fig. \ref{attochirpscans_shorterth}b), which, in addition, agree with the results presented in a recent publication \cite{timmers2017postcompression}, showing that similar duration driving pulses can {directly} generate {isolated} attosecond pulses {via intensity gating of the} HHG process. Again, as seen in the corresponding spectra from the HHG dispersion scans, high HHG yield in the achromatic case is much more restricted to a low range of material insertion. In both cases, the single attosecond pulses present a duration of around 500 as (FWHM). The effect of spectral phase on the HHG dispersion scan can be observed by comparing Fig. \ref{attochirpscans_shorterth}a and Fig. \ref{attochirpscans_shorterth}b with the corresponding {temporal} structure {obtained for a} 2.9 FWHM FL driving pulse (Fig. \ref{attochirpscans_shorterth}c). In the {latter}, single attosecond pulses can be observed exhibiting shorter duration (around 400 as), compatible with the fact that {the} HHG spectrum presents {a} higher cut-off (Fig. \ref{attochirpscans_shorterth}c) than {in} the other studied cases. Therefore, according to the simulations, the chromatic focusing by means of a singlet {lens} does not disrupt the single attosecond pulse generation at the considered input pulse characteristics, allowing the generation of a single attosecond burst when near single-cycle driving pulses are used.}

	\section{Conclusions}
    {{In conclusion}, we have shown that focusing ultra-broadband pulses with some chromatic dependence (e.g., with a singlet lens) can provide {an additional} degree of spectral control {over} the resulting HHG spectrum, while no major distortions disrupting the HHG process are introduced. The focusing of each wavelength component at different focal lengths has an impact in the single atom emission across the generating region and also in the coherent addition of the XUV light from all the elemental emitters. Therefore, an additional experimental parameter, the chromatic focal distribution, can be used for altering the XUV characteristics, namely {for HHG} spectral tuning. To {demonstrate this effect}, we have used a singlet {lens, which introduces chromatic aberration,} with broadband few-cycle pulses. We have observed that it is possible to tune the harmonics by changing the singlet position when focusing few-cycle pulses, since the spectral content {at} the focus changes {due to} chromatic aberration. This ability can be useful to perform spectroscopy in the XUV region. On the other hand, the employed focusing scheme does not modify some of the properties observed in HHG using achromatic focusing, namely {the possibility to} generate spectral XUV continua even in the low-order harmonic region under adequate pulse (here shown from 20 eV to 39 eV) and phase matching conditions, or even the generation of {an isolated} single attosecond pulse {from} near single-cycle {driving} pulses. Furthermore, {the chromatic focusing scheme} produces XUV radiation with a similar yield {to the achromatic case}.}
    
    {Full-propagation theoretical simulations {corroborate} the experimental results, confirming that the chromatic focusing reproduces the temporal structure obtained in the achromatic case. Furthermore, since the HHG process is distributed in different spatial regions depending on the wavelength, further control over the chromatic focusing (e.g., using diffractive lenses, adaptive optics, etc.) and the gas distribution profile may allow to tailor the phase-matching conditions with an additional control parameter offering further optimization of the HHG process.}
	
	\vspace{0.8cm}
	
	\section{ACKNOWLEDGMENTS}
	
	We acknowledge support from Junta de Castilla y León (Projects No. SA116U13 and SA046U16), Spanish MINECO (Grants No. FIS2009-09522, No. FIS2013-44174 P, No. FIS2015-71933-REDT and No. FIS2016-75652-P). This work was partly supported by Fundação para a Ciência e Tecnologia, Portugal, co-funded by COMPETE and FEDER, via Grants PTDC/FIS/122511/2010 and UID/NAN/50024/2013; F.S. acknowledges support from Grant SFRH/BD/69913/2010 and B.A. acknowledges support from Post-Doctoral Fellowship SFRH/BPD/88424/2012; H.C. acknowledges support from Sabbatical Leave Grant SFRH/BSAB/105974/2015. C.H-G acknowledges support from a Marie Curie International Outgoing Fellowship within the EU Seventh Framework Programme for Research and Technological Development, under REA Grant Agreement No. 328334.

	\bibliography{CromaticaBiblio}

\begin{thebibliography}{44}%
\makeatletter
\providecommand \@ifxundefined [1]{%
 \@ifx{#1\undefined}
}%
\providecommand \@ifnum [1]{%
 \ifnum #1\expandafter \@firstoftwo
 \else \expandafter \@secondoftwo
 \fi
}%
\providecommand \@ifx [1]{%
 \ifx #1\expandafter \@firstoftwo
 \else \expandafter \@secondoftwo
 \fi
}%
\providecommand \natexlab [1]{#1}%
\providecommand \enquote  [1]{``#1''}%
\providecommand \bibnamefont  [1]{#1}%
\providecommand \bibfnamefont [1]{#1}%
\providecommand \citenamefont [1]{#1}%
\providecommand \href@noop [0]{\@secondoftwo}%
\providecommand \href [0]{\begingroup \@sanitize@url \@href}%
\providecommand \@href[1]{\@@startlink{#1}\@@href}%
\providecommand \@@href[1]{\endgroup#1\@@endlink}%
\providecommand \@sanitize@url [0]{\catcode `\\12\catcode `\$12\catcode
  `\&12\catcode `\#12\catcode `\^12\catcode `\_12\catcode `\%12\relax}%
\providecommand \@@startlink[1]{}%
\providecommand \@@endlink[0]{}%
\providecommand \url  [0]{\begingroup\@sanitize@url \@url }%
\providecommand \@url [1]{\endgroup\@href {#1}{\urlprefix }}%
\providecommand \urlprefix  [0]{URL }%
\providecommand \Eprint [0]{\href }%
\providecommand \doibase [0]{http://dx.doi.org/}%
\providecommand \selectlanguage [0]{\@gobble}%
\providecommand \bibinfo  [0]{\@secondoftwo}%
\providecommand \bibfield  [0]{\@secondoftwo}%
\providecommand \translation [1]{[#1]}%
\providecommand \BibitemOpen [0]{}%
\providecommand \bibitemStop [0]{}%
\providecommand \bibitemNoStop [0]{.\EOS\space}%
\providecommand \EOS [0]{\spacefactor3000\relax}%
\providecommand \BibitemShut  [1]{\csname bibitem#1\endcsname}%
\let\auto@bib@innerbib\@empty
\bibitem [{\citenamefont {Hentschel}\ \emph {et~al.}(2001)\citenamefont
  {Hentschel}, \citenamefont {Kienberger}, \citenamefont {Spielmann},
  \citenamefont {Reider}, \citenamefont {Milosevic}, \citenamefont {Brabec},
  \citenamefont {Corkum}, \citenamefont {Heinzmann}, \citenamefont {Drescher},\
  and\ \citenamefont {Krausz}}]{hentschel2001attosecond}%
  \BibitemOpen
  \bibfield  {author} {\bibinfo {author} {\bibfnamefont {M.}~\bibnamefont
  {Hentschel}}, \bibinfo {author} {\bibfnamefont {R.}~\bibnamefont
  {Kienberger}}, \bibinfo {author} {\bibfnamefont {C.}~\bibnamefont
  {Spielmann}}, \bibinfo {author} {\bibfnamefont {G.~A.}\ \bibnamefont
  {Reider}}, \bibinfo {author} {\bibfnamefont {N.}~\bibnamefont {Milosevic}},
  \bibinfo {author} {\bibfnamefont {T.}~\bibnamefont {Brabec}}, \bibinfo
  {author} {\bibfnamefont {P.~B.}\ \bibnamefont {Corkum}}, \bibinfo {author}
  {\bibfnamefont {U.}~\bibnamefont {Heinzmann}}, \bibinfo {author}
  {\bibfnamefont {M.}~\bibnamefont {Drescher}}, \ and\ \bibinfo {author}
  {\bibfnamefont {F.}~\bibnamefont {Krausz}},\ }\href@noop {} {\bibfield
  {journal} {\bibinfo  {journal} {Nature}\ }\textbf {\bibinfo {volume} {414}},\
  \bibinfo {pages} {509} (\bibinfo {year} {2001})}\BibitemShut {NoStop}%
\bibitem [{\citenamefont {Sansone}\ \emph {et~al.}(2006)\citenamefont
  {Sansone}, \citenamefont {Benedetti}, \citenamefont {Calegari}, \citenamefont
  {Vozzi}, \citenamefont {Avaldi}, \citenamefont {Flammini}, \citenamefont
  {Poletto}, \citenamefont {Villoresi}, \citenamefont {Altucci}, \citenamefont
  {Velotta}, \citenamefont {Stagira}, \citenamefont {Silvestri},\ and\
  \citenamefont {Nisoli}}]{sansone2006isolated}%
  \BibitemOpen
  \bibfield  {author} {\bibinfo {author} {\bibfnamefont {G.}~\bibnamefont
  {Sansone}}, \bibinfo {author} {\bibfnamefont {E.}~\bibnamefont {Benedetti}},
  \bibinfo {author} {\bibfnamefont {F.}~\bibnamefont {Calegari}}, \bibinfo
  {author} {\bibfnamefont {C.}~\bibnamefont {Vozzi}}, \bibinfo {author}
  {\bibfnamefont {L.}~\bibnamefont {Avaldi}}, \bibinfo {author} {\bibfnamefont
  {R.}~\bibnamefont {Flammini}}, \bibinfo {author} {\bibfnamefont
  {L.}~\bibnamefont {Poletto}}, \bibinfo {author} {\bibfnamefont
  {P.}~\bibnamefont {Villoresi}}, \bibinfo {author} {\bibfnamefont
  {C.}~\bibnamefont {Altucci}}, \bibinfo {author} {\bibfnamefont
  {R.}~\bibnamefont {Velotta}}, \bibinfo {author} {\bibfnamefont
  {S.}~\bibnamefont {Stagira}}, \bibinfo {author} {\bibfnamefont {S.~D.}\
  \bibnamefont {Silvestri}}, \ and\ \bibinfo {author} {\bibfnamefont
  {M.}~\bibnamefont {Nisoli}},\ }\href@noop {} {\bibfield  {journal} {\bibinfo
  {journal} {Science}\ }\textbf {\bibinfo {volume} {314}},\ \bibinfo {pages}
  {443} (\bibinfo {year} {2006})}\BibitemShut {NoStop}%
\bibitem [{\citenamefont {Corkum}\ and\ \citenamefont
  {Krausz}(2007)}]{corkum2007attosecond}%
  \BibitemOpen
  \bibfield  {author} {\bibinfo {author} {\bibfnamefont {P.~B.}\ \bibnamefont
  {Corkum}}\ and\ \bibinfo {author} {\bibfnamefont {F.}~\bibnamefont
  {Krausz}},\ }\href@noop {} {\bibfield  {journal} {\bibinfo  {journal} {Nat.
  Phys.}\ }\textbf {\bibinfo {volume} {3}},\ \bibinfo {pages} {381} (\bibinfo
  {year} {2007})}\BibitemShut {NoStop}%
\bibitem [{\citenamefont {Goulielmakis}\ \emph {et~al.}(2010)\citenamefont
  {Goulielmakis}, \citenamefont {Loh}, \citenamefont {Wirth}, \citenamefont
  {Santra}, \citenamefont {Rohringer}, \citenamefont {Yakovlev}, \citenamefont
  {Zherebtsov}, \citenamefont {Pfeifer}, \citenamefont {Azzeer}, \citenamefont
  {Kling}, \citenamefont {Leone},\ and\ \citenamefont
  {Krausz}}]{goulielmakis2010real}%
  \BibitemOpen
  \bibfield  {author} {\bibinfo {author} {\bibfnamefont {E.}~\bibnamefont
  {Goulielmakis}}, \bibinfo {author} {\bibfnamefont {Z.-H.}\ \bibnamefont
  {Loh}}, \bibinfo {author} {\bibfnamefont {A.}~\bibnamefont {Wirth}}, \bibinfo
  {author} {\bibfnamefont {R.}~\bibnamefont {Santra}}, \bibinfo {author}
  {\bibfnamefont {N.}~\bibnamefont {Rohringer}}, \bibinfo {author}
  {\bibfnamefont {V.~S.}\ \bibnamefont {Yakovlev}}, \bibinfo {author}
  {\bibfnamefont {S.}~\bibnamefont {Zherebtsov}}, \bibinfo {author}
  {\bibfnamefont {T.}~\bibnamefont {Pfeifer}}, \bibinfo {author} {\bibfnamefont
  {A.~M.}\ \bibnamefont {Azzeer}}, \bibinfo {author} {\bibfnamefont {M.~F.}\
  \bibnamefont {Kling}}, \bibinfo {author} {\bibfnamefont {S.~R.}\ \bibnamefont
  {Leone}}, \ and\ \bibinfo {author} {\bibfnamefont {F.}~\bibnamefont
  {Krausz}},\ }\href@noop {} {\bibfield  {journal} {\bibinfo  {journal}
  {Nature}\ }\textbf {\bibinfo {volume} {466}},\ \bibinfo {pages} {739}
  (\bibinfo {year} {2010})}\BibitemShut {NoStop}%
\bibitem [{\citenamefont {Fuchs}\ \emph {et~al.}(2012)\citenamefont {Fuchs},
  \citenamefont {Blinne}, \citenamefont {R{\"o}del}, \citenamefont {Zastrau},
  \citenamefont {Hilbert}, \citenamefont {W{\"u}nsche}, \citenamefont
  {Bierbach}, \citenamefont {Frumker}, \citenamefont {F{\"o}rster},\ and\
  \citenamefont {Paulus}}]{fuchs2012optical}%
  \BibitemOpen
  \bibfield  {author} {\bibinfo {author} {\bibfnamefont {S.}~\bibnamefont
  {Fuchs}}, \bibinfo {author} {\bibfnamefont {A.}~\bibnamefont {Blinne}},
  \bibinfo {author} {\bibfnamefont {C.}~\bibnamefont {R{\"o}del}}, \bibinfo
  {author} {\bibfnamefont {U.}~\bibnamefont {Zastrau}}, \bibinfo {author}
  {\bibfnamefont {V.}~\bibnamefont {Hilbert}}, \bibinfo {author} {\bibfnamefont
  {M.}~\bibnamefont {W{\"u}nsche}}, \bibinfo {author} {\bibfnamefont
  {J.}~\bibnamefont {Bierbach}}, \bibinfo {author} {\bibfnamefont
  {E.}~\bibnamefont {Frumker}}, \bibinfo {author} {\bibfnamefont
  {E.}~\bibnamefont {F{\"o}rster}}, \ and\ \bibinfo {author} {\bibfnamefont
  {G.~G.}\ \bibnamefont {Paulus}},\ }\href@noop {} {\bibfield  {journal}
  {\bibinfo  {journal} {Appl. Phys. B}\ }\textbf {\bibinfo {volume} {106}},\
  \bibinfo {pages} {789} (\bibinfo {year} {2012})}\BibitemShut {NoStop}%
\bibitem [{\citenamefont {Nisoli}, \citenamefont {De~Silvestri},\ and\
  \citenamefont {Svelto}(1996)}]{nisoli1996generation}%
  \BibitemOpen
  \bibfield  {author} {\bibinfo {author} {\bibfnamefont {M.}~\bibnamefont
  {Nisoli}}, \bibinfo {author} {\bibfnamefont {S.}~\bibnamefont
  {De~Silvestri}}, \ and\ \bibinfo {author} {\bibfnamefont {O.}~\bibnamefont
  {Svelto}},\ }\href@noop {} {\bibfield  {journal} {\bibinfo  {journal} {Appl.
  Phys. Lett.}\ }\textbf {\bibinfo {volume} {68}},\ \bibinfo {pages} {2793}
  (\bibinfo {year} {1996})}\BibitemShut {NoStop}%
\bibitem [{\citenamefont {Sartania}\ \emph {et~al.}(1997)\citenamefont
  {Sartania}, \citenamefont {Cheng}, \citenamefont {Lenzner}, \citenamefont
  {Tempea}, \citenamefont {Spielmann}, \citenamefont {Krausz},\ and\
  \citenamefont {Ferencz}}]{Sartania:97}%
  \BibitemOpen
  \bibfield  {author} {\bibinfo {author} {\bibfnamefont {S.}~\bibnamefont
  {Sartania}}, \bibinfo {author} {\bibfnamefont {Z.}~\bibnamefont {Cheng}},
  \bibinfo {author} {\bibfnamefont {M.}~\bibnamefont {Lenzner}}, \bibinfo
  {author} {\bibfnamefont {G.}~\bibnamefont {Tempea}}, \bibinfo {author}
  {\bibfnamefont {C.}~\bibnamefont {Spielmann}}, \bibinfo {author}
  {\bibfnamefont {F.}~\bibnamefont {Krausz}}, \ and\ \bibinfo {author}
  {\bibfnamefont {K.}~\bibnamefont {Ferencz}},\ }\href@noop {} {\bibfield
  {journal} {\bibinfo  {journal} {Opt. Lett.}\ }\textbf {\bibinfo {volume}
  {22}},\ \bibinfo {pages} {1562} (\bibinfo {year} {1997})}\BibitemShut
  {NoStop}%
\bibitem [{\citenamefont {Hauri}\ \emph {et~al.}(2004)\citenamefont {Hauri},
  \citenamefont {Kornelis}, \citenamefont {Helbing}, \citenamefont {Heinrich},
  \citenamefont {Couairon}, \citenamefont {Mysyrowicz}, \citenamefont
  {Biegert},\ and\ \citenamefont {Keller}}]{hauri2004generation}%
  \BibitemOpen
  \bibfield  {author} {\bibinfo {author} {\bibfnamefont {C.}~\bibnamefont
  {Hauri}}, \bibinfo {author} {\bibfnamefont {W.}~\bibnamefont {Kornelis}},
  \bibinfo {author} {\bibfnamefont {F.}~\bibnamefont {Helbing}}, \bibinfo
  {author} {\bibfnamefont {A.}~\bibnamefont {Heinrich}}, \bibinfo {author}
  {\bibfnamefont {A.}~\bibnamefont {Couairon}}, \bibinfo {author}
  {\bibfnamefont {A.}~\bibnamefont {Mysyrowicz}}, \bibinfo {author}
  {\bibfnamefont {J.}~\bibnamefont {Biegert}}, \ and\ \bibinfo {author}
  {\bibfnamefont {U.}~\bibnamefont {Keller}},\ }\href@noop {} {\bibfield
  {journal} {\bibinfo  {journal} {Appl. Phys. B}\ }\textbf {\bibinfo {volume}
  {79}},\ \bibinfo {pages} {673} (\bibinfo {year} {2004})}\BibitemShut
  {NoStop}%
\bibitem [{\citenamefont {Silva}\ \emph {et~al.}(2014)\citenamefont {Silva},
  \citenamefont {Miranda}, \citenamefont {Alonso}, \citenamefont
  {Rauschenberger}, \citenamefont {Pervak},\ and\ \citenamefont
  {Crespo}}]{silva2014simultaneous}%
  \BibitemOpen
  \bibfield  {author} {\bibinfo {author} {\bibfnamefont {F.}~\bibnamefont
  {Silva}}, \bibinfo {author} {\bibfnamefont {M.}~\bibnamefont {Miranda}},
  \bibinfo {author} {\bibfnamefont {B.}~\bibnamefont {Alonso}}, \bibinfo
  {author} {\bibfnamefont {J.}~\bibnamefont {Rauschenberger}}, \bibinfo
  {author} {\bibfnamefont {V.}~\bibnamefont {Pervak}}, \ and\ \bibinfo {author}
  {\bibfnamefont {H.}~\bibnamefont {Crespo}},\ }\href@noop {} {\bibfield
  {journal} {\bibinfo  {journal} {Opt. Express}\ }\textbf {\bibinfo {volume}
  {22}},\ \bibinfo {pages} {10181} (\bibinfo {year} {2014})}\BibitemShut
  {NoStop}%
\bibitem [{\citenamefont {Wirth}\ \emph {et~al.}(2011)\citenamefont {Wirth},
  \citenamefont {Hassan}, \citenamefont {Grgura{\v{s}}}, \citenamefont
  {Gagnon}, \citenamefont {Moulet}, \citenamefont {Luu}, \citenamefont {Pabst},
  \citenamefont {Santra}, \citenamefont {Alahmed}, \citenamefont {Azzeer},
  \citenamefont {Yakovlev}, \citenamefont {Pervak}, \citenamefont {Krausz},\
  and\ \citenamefont {Goulielmakis}}]{wirth2011synthesized}%
  \BibitemOpen
  \bibfield  {author} {\bibinfo {author} {\bibfnamefont {A.}~\bibnamefont
  {Wirth}}, \bibinfo {author} {\bibfnamefont {M.~T.}\ \bibnamefont {Hassan}},
  \bibinfo {author} {\bibfnamefont {I.}~\bibnamefont {Grgura{\v{s}}}}, \bibinfo
  {author} {\bibfnamefont {J.}~\bibnamefont {Gagnon}}, \bibinfo {author}
  {\bibfnamefont {A.}~\bibnamefont {Moulet}}, \bibinfo {author} {\bibfnamefont
  {T.~T.}\ \bibnamefont {Luu}}, \bibinfo {author} {\bibfnamefont
  {S.}~\bibnamefont {Pabst}}, \bibinfo {author} {\bibfnamefont
  {R.}~\bibnamefont {Santra}}, \bibinfo {author} {\bibfnamefont
  {Z.}~\bibnamefont {Alahmed}}, \bibinfo {author} {\bibfnamefont
  {A.}~\bibnamefont {Azzeer}}, \bibinfo {author} {\bibfnamefont {V.~S.}\
  \bibnamefont {Yakovlev}}, \bibinfo {author} {\bibfnamefont {V.}~\bibnamefont
  {Pervak}}, \bibinfo {author} {\bibfnamefont {F.}~\bibnamefont {Krausz}}, \
  and\ \bibinfo {author} {\bibfnamefont {E.}~\bibnamefont {Goulielmakis}},\
  }\href@noop {} {\bibfield  {journal} {\bibinfo  {journal} {Science}\ }\textbf
  {\bibinfo {volume} {334}},\ \bibinfo {pages} {195} (\bibinfo {year}
  {2011})}\BibitemShut {NoStop}%
\bibitem [{\citenamefont {Alonso}\ \emph
  {et~al.}(2013{\natexlab{a}})\citenamefont {Alonso}, \citenamefont {Miranda},
  \citenamefont {Sola},\ and\ \citenamefont
  {Crespo}}]{alonso2013spatiotemporal}%
  \BibitemOpen
  \bibfield  {author} {\bibinfo {author} {\bibfnamefont {B.}~\bibnamefont
  {Alonso}}, \bibinfo {author} {\bibfnamefont {M.}~\bibnamefont {Miranda}},
  \bibinfo {author} {\bibfnamefont {{\'I}.~J.}\ \bibnamefont {Sola}}, \ and\
  \bibinfo {author} {\bibfnamefont {H.}~\bibnamefont {Crespo}},\ }\href@noop {}
  {\bibfield  {journal} {\bibinfo  {journal} {Opt. Express}\ }\textbf {\bibinfo
  {volume} {21}},\ \bibinfo {pages} {5582} (\bibinfo {year}
  {2013}{\natexlab{a}})}\BibitemShut {NoStop}%
\bibitem [{\citenamefont {Alonso}\ \emph
  {et~al.}(2011{\natexlab{a}})\citenamefont {Alonso}, \citenamefont {Sola},
  \citenamefont {San~Rom{\'a}n}, \citenamefont {Varela},\ and\ \citenamefont
  {Roso}}]{alonso2011spatiotemporal}%
  \BibitemOpen
  \bibfield  {author} {\bibinfo {author} {\bibfnamefont {B.}~\bibnamefont
  {Alonso}}, \bibinfo {author} {\bibfnamefont {{\'I}.~J.}\ \bibnamefont
  {Sola}}, \bibinfo {author} {\bibfnamefont {J.}~\bibnamefont {San~Rom{\'a}n}},
  \bibinfo {author} {\bibfnamefont {{\'O}.}~\bibnamefont {Varela}}, \ and\
  \bibinfo {author} {\bibfnamefont {L.}~\bibnamefont {Roso}},\ }\href@noop {}
  {\bibfield  {journal} {\bibinfo  {journal} {J. Opt. Soc. Am. B}\ }\textbf
  {\bibinfo {volume} {28}},\ \bibinfo {pages} {1807} (\bibinfo {year}
  {2011}{\natexlab{a}})}\BibitemShut {NoStop}%
\bibitem [{\citenamefont {Alonso}\ \emph
  {et~al.}(2013{\natexlab{b}})\citenamefont {Alonso}, \citenamefont {Miranda},
  \citenamefont {Silva}, \citenamefont {Pervak}, \citenamefont
  {Rauschenberger}, \citenamefont {San~Rom{\'a}n}, \citenamefont {Sola},\ and\
  \citenamefont {Crespo}}]{alonso2013characterization}%
  \BibitemOpen
  \bibfield  {author} {\bibinfo {author} {\bibfnamefont {B.}~\bibnamefont
  {Alonso}}, \bibinfo {author} {\bibfnamefont {M.}~\bibnamefont {Miranda}},
  \bibinfo {author} {\bibfnamefont {F.}~\bibnamefont {Silva}}, \bibinfo
  {author} {\bibfnamefont {V.}~\bibnamefont {Pervak}}, \bibinfo {author}
  {\bibfnamefont {J.}~\bibnamefont {Rauschenberger}}, \bibinfo {author}
  {\bibfnamefont {J.}~\bibnamefont {San~Rom{\'a}n}}, \bibinfo {author}
  {\bibfnamefont {{\'I}.~J.}\ \bibnamefont {Sola}}, \ and\ \bibinfo {author}
  {\bibfnamefont {H.}~\bibnamefont {Crespo}},\ }\href@noop {} {\bibfield
  {journal} {\bibinfo  {journal} {Appl. Phys. B}\ }\textbf {\bibinfo {volume}
  {112}},\ \bibinfo {pages} {105} (\bibinfo {year}
  {2013}{\natexlab{b}})}\BibitemShut {NoStop}%
\bibitem [{\citenamefont {Alonso}\ \emph {et~al.}(2012)\citenamefont {Alonso},
  \citenamefont {Borrego-Varillas}, \citenamefont {Mendoza-Yero}, \citenamefont
  {Sola}, \citenamefont {San~Rom{\'a}n}, \citenamefont {M{\'\i}nguez-Vega},\
  and\ \citenamefont {Roso}}]{alonso2012frequency}%
  \BibitemOpen
  \bibfield  {author} {\bibinfo {author} {\bibfnamefont {B.}~\bibnamefont
  {Alonso}}, \bibinfo {author} {\bibfnamefont {R.}~\bibnamefont
  {Borrego-Varillas}}, \bibinfo {author} {\bibfnamefont {O.}~\bibnamefont
  {Mendoza-Yero}}, \bibinfo {author} {\bibfnamefont {{\'I}.~J.}\ \bibnamefont
  {Sola}}, \bibinfo {author} {\bibfnamefont {J.}~\bibnamefont {San~Rom{\'a}n}},
  \bibinfo {author} {\bibfnamefont {G.}~\bibnamefont {M{\'\i}nguez-Vega}}, \
  and\ \bibinfo {author} {\bibfnamefont {L.}~\bibnamefont {Roso}},\ }\href@noop
  {} {\bibfield  {journal} {\bibinfo  {journal} {J. Opt. Soc. Am. B}\ }\textbf
  {\bibinfo {volume} {29}},\ \bibinfo {pages} {1993} (\bibinfo {year}
  {2012})}\BibitemShut {NoStop}%
\bibitem [{\citenamefont {M{\'\i}nguez-Vega}\ \emph {et~al.}(2010)\citenamefont
  {M{\'\i}nguez-Vega}, \citenamefont {Romero}, \citenamefont {Mendoza-Yero},
  \citenamefont {de~Aldana}, \citenamefont {Borrego-Varillas}, \citenamefont
  {M{\'e}ndez}, \citenamefont {Andr{\'e}s}, \citenamefont {Lancis},
  \citenamefont {Climent},\ and\ \citenamefont {Roso}}]{minguez2010wavelength}%
  \BibitemOpen
  \bibfield  {author} {\bibinfo {author} {\bibfnamefont {G.}~\bibnamefont
  {M{\'\i}nguez-Vega}}, \bibinfo {author} {\bibfnamefont {C.}~\bibnamefont
  {Romero}}, \bibinfo {author} {\bibfnamefont {O.}~\bibnamefont
  {Mendoza-Yero}}, \bibinfo {author} {\bibfnamefont {J.~V.}\ \bibnamefont
  {de~Aldana}}, \bibinfo {author} {\bibfnamefont {R.}~\bibnamefont
  {Borrego-Varillas}}, \bibinfo {author} {\bibfnamefont {C.}~\bibnamefont
  {M{\'e}ndez}}, \bibinfo {author} {\bibfnamefont {P.}~\bibnamefont
  {Andr{\'e}s}}, \bibinfo {author} {\bibfnamefont {J.}~\bibnamefont {Lancis}},
  \bibinfo {author} {\bibfnamefont {V.}~\bibnamefont {Climent}}, \ and\
  \bibinfo {author} {\bibfnamefont {L.}~\bibnamefont {Roso}},\ }\href@noop {}
  {\bibfield  {journal} {\bibinfo  {journal} {Opt. Lett.}\ }\textbf {\bibinfo
  {volume} {35}},\ \bibinfo {pages} {3694} (\bibinfo {year}
  {2010})}\BibitemShut {NoStop}%
\bibitem [{\citenamefont {Borrego-Varillas}\ \emph {et~al.}(2013)\citenamefont
  {Borrego-Varillas}, \citenamefont {Romero}, \citenamefont {Mendoza-Yero},
  \citenamefont {M{\'\i}nguez-Vega}, \citenamefont {Gallardo},\ and\
  \citenamefont {de~Aldana}}]{borrego2013femtosecond}%
  \BibitemOpen
  \bibfield  {author} {\bibinfo {author} {\bibfnamefont {R.}~\bibnamefont
  {Borrego-Varillas}}, \bibinfo {author} {\bibfnamefont {C.}~\bibnamefont
  {Romero}}, \bibinfo {author} {\bibfnamefont {O.}~\bibnamefont
  {Mendoza-Yero}}, \bibinfo {author} {\bibfnamefont {G.}~\bibnamefont
  {M{\'\i}nguez-Vega}}, \bibinfo {author} {\bibfnamefont {I.}~\bibnamefont
  {Gallardo}}, \ and\ \bibinfo {author} {\bibfnamefont {J.~V.}\ \bibnamefont
  {de~Aldana}},\ }\href@noop {} {\bibfield  {journal} {\bibinfo  {journal} {J.
  Opt. Soc. Am. B}\ }\textbf {\bibinfo {volume} {30}},\ \bibinfo {pages} {2059}
  (\bibinfo {year} {2013})}\BibitemShut {NoStop}%
\bibitem [{\citenamefont {Alonso}\ \emph
  {et~al.}(2011{\natexlab{b}})\citenamefont {Alonso}, \citenamefont
  {Borrego-Varillas}, \citenamefont {Sola}, \citenamefont {Varela},
  \citenamefont {Villamar{\'\i}n}, \citenamefont {Collados}, \citenamefont
  {San~Rom{\'a}n}, \citenamefont {Bueno},\ and\ \citenamefont
  {Roso}}]{alonso2011enhancement}%
  \BibitemOpen
  \bibfield  {author} {\bibinfo {author} {\bibfnamefont {B.}~\bibnamefont
  {Alonso}}, \bibinfo {author} {\bibfnamefont {R.}~\bibnamefont
  {Borrego-Varillas}}, \bibinfo {author} {\bibfnamefont {{\'I}.~J.}\
  \bibnamefont {Sola}}, \bibinfo {author} {\bibfnamefont {{\'O}.}~\bibnamefont
  {Varela}}, \bibinfo {author} {\bibfnamefont {A.}~\bibnamefont
  {Villamar{\'\i}n}}, \bibinfo {author} {\bibfnamefont {M.~V.}\ \bibnamefont
  {Collados}}, \bibinfo {author} {\bibfnamefont {J.}~\bibnamefont
  {San~Rom{\'a}n}}, \bibinfo {author} {\bibfnamefont {J.~M.}\ \bibnamefont
  {Bueno}}, \ and\ \bibinfo {author} {\bibfnamefont {L.}~\bibnamefont {Roso}},\
  }\href@noop {} {\bibfield  {journal} {\bibinfo  {journal} {Opt. Lett.}\
  }\textbf {\bibinfo {volume} {36}},\ \bibinfo {pages} {3867} (\bibinfo {year}
  {2011}{\natexlab{b}})}\BibitemShut {NoStop}%
\bibitem [{\citenamefont {Gualda}, \citenamefont {Bueno},\ and\ \citenamefont
  {Artal}(2010)}]{gualda2010wavefront}%
  \BibitemOpen
  \bibfield  {author} {\bibinfo {author} {\bibfnamefont {E.~J.}\ \bibnamefont
  {Gualda}}, \bibinfo {author} {\bibfnamefont {J.~M.}\ \bibnamefont {Bueno}}, \
  and\ \bibinfo {author} {\bibfnamefont {P.}~\bibnamefont {Artal}},\
  }\href@noop {} {\bibfield  {journal} {\bibinfo  {journal} {J. of Biomedical
  Optics}\ }\textbf {\bibinfo {volume} {15}},\ \bibinfo {pages} {026007}
  (\bibinfo {year} {2010})}\BibitemShut {NoStop}%
\bibitem [{\citenamefont {Wheeler}\ \emph {et~al.}(2012)\citenamefont
  {Wheeler}, \citenamefont {Borot}, \citenamefont {Monchoc{\'e}}, \citenamefont
  {Vincenti}, \citenamefont {Ricci}, \citenamefont {Malvache}, \citenamefont
  {Lopez-Martens},\ and\ \citenamefont
  {Qu{\'e}r{\'e}}}]{wheeler2012attosecond}%
  \BibitemOpen
  \bibfield  {author} {\bibinfo {author} {\bibfnamefont {J.~A.}\ \bibnamefont
  {Wheeler}}, \bibinfo {author} {\bibfnamefont {A.}~\bibnamefont {Borot}},
  \bibinfo {author} {\bibfnamefont {S.}~\bibnamefont {Monchoc{\'e}}}, \bibinfo
  {author} {\bibfnamefont {H.}~\bibnamefont {Vincenti}}, \bibinfo {author}
  {\bibfnamefont {A.}~\bibnamefont {Ricci}}, \bibinfo {author} {\bibfnamefont
  {A.}~\bibnamefont {Malvache}}, \bibinfo {author} {\bibfnamefont
  {R.}~\bibnamefont {Lopez-Martens}}, \ and\ \bibinfo {author} {\bibfnamefont
  {F.}~\bibnamefont {Qu{\'e}r{\'e}}},\ }\href@noop {} {\bibfield  {journal}
  {\bibinfo  {journal} {Nat. Photon.}\ }\textbf {\bibinfo {volume} {6}},\
  \bibinfo {pages} {829} (\bibinfo {year} {2012})}\BibitemShut {NoStop}%
\bibitem [{\citenamefont {Hern\'andez-Garc\'ia}\ \emph
  {et~al.}(2016)\citenamefont {Hern\'andez-Garc\'ia}, \citenamefont
  {Jaron-Becker}, \citenamefont {Hickstein}, \citenamefont {Becker},\ and\
  \citenamefont {Durfee}}]{hernandez2016high}%
  \BibitemOpen
  \bibfield  {author} {\bibinfo {author} {\bibfnamefont {C.}~\bibnamefont
  {Hern\'andez-Garc\'ia}}, \bibinfo {author} {\bibfnamefont {A.}~\bibnamefont
  {Jaron-Becker}}, \bibinfo {author} {\bibfnamefont {D.~D.}\ \bibnamefont
  {Hickstein}}, \bibinfo {author} {\bibfnamefont {A.}~\bibnamefont {Becker}}, \
  and\ \bibinfo {author} {\bibfnamefont {C.~G.}\ \bibnamefont {Durfee}},\
  }\href@noop {} {\bibfield  {journal} {\bibinfo  {journal} {Phys. Rev. A}\
  }\textbf {\bibinfo {volume} {93}},\ \bibinfo {pages} {023825} (\bibinfo
  {year} {2016})}\BibitemShut {NoStop}%
\bibitem [{\citenamefont {Shan}, \citenamefont {Cavalieri},\ and\ \citenamefont
  {Chang}(2002)}]{shan2002tunable}%
  \BibitemOpen
  \bibfield  {author} {\bibinfo {author} {\bibfnamefont {B.}~\bibnamefont
  {Shan}}, \bibinfo {author} {\bibfnamefont {A.}~\bibnamefont {Cavalieri}}, \
  and\ \bibinfo {author} {\bibfnamefont {Z.}~\bibnamefont {Chang}},\
  }\href@noop {} {\bibfield  {journal} {\bibinfo  {journal} {Appl. Phys. B}\
  }\textbf {\bibinfo {volume} {74}},\ \bibinfo {pages} {s23} (\bibinfo {year}
  {2002})}\BibitemShut {NoStop}%
\bibitem [{\citenamefont {Reitze}\ \emph {et~al.}(2004)\citenamefont {Reitze},
  \citenamefont {Kazamias}, \citenamefont {Weihe}, \citenamefont {Mullot},
  \citenamefont {Douillet}, \citenamefont {Aug{\'e}}, \citenamefont {Albert},
  \citenamefont {Ramanathan}, \citenamefont {Chambaret}, \citenamefont {Hulin}
  \emph {et~al.}}]{reitze2004enhancement}%
  \BibitemOpen
  \bibfield  {author} {\bibinfo {author} {\bibfnamefont {D.~H.}\ \bibnamefont
  {Reitze}}, \bibinfo {author} {\bibfnamefont {S.}~\bibnamefont {Kazamias}},
  \bibinfo {author} {\bibfnamefont {F.}~\bibnamefont {Weihe}}, \bibinfo
  {author} {\bibfnamefont {G.}~\bibnamefont {Mullot}}, \bibinfo {author}
  {\bibfnamefont {D.}~\bibnamefont {Douillet}}, \bibinfo {author}
  {\bibfnamefont {F.}~\bibnamefont {Aug{\'e}}}, \bibinfo {author}
  {\bibfnamefont {O.}~\bibnamefont {Albert}}, \bibinfo {author} {\bibfnamefont
  {V.}~\bibnamefont {Ramanathan}}, \bibinfo {author} {\bibfnamefont {J.~P.}\
  \bibnamefont {Chambaret}}, \bibinfo {author} {\bibfnamefont {D.}~\bibnamefont
  {Hulin}},  \emph {et~al.},\ }\href@noop {} {\bibfield  {journal} {\bibinfo
  {journal} {Opt. Lett.}\ }\textbf {\bibinfo {volume} {29}},\ \bibinfo {pages}
  {86} (\bibinfo {year} {2004})}\BibitemShut {NoStop}%
\bibitem [{\citenamefont {Lu}\ \emph {et~al.}(2013)\citenamefont {Lu},
  \citenamefont {Zhang}, \citenamefont {Xia},\ and\ \citenamefont
  {Chen}}]{lu2013generation}%
  \BibitemOpen
  \bibfield  {author} {\bibinfo {author} {\bibfnamefont {F.}~\bibnamefont
  {Lu}}, \bibinfo {author} {\bibfnamefont {S.}~\bibnamefont {Zhang}}, \bibinfo
  {author} {\bibfnamefont {Y.}~\bibnamefont {Xia}}, \ and\ \bibinfo {author}
  {\bibfnamefont {D.}~\bibnamefont {Chen}},\ }\href@noop {} {\bibfield
  {journal} {\bibinfo  {journal} {Laser Phys.}\ }\textbf {\bibinfo {volume}
  {23}},\ \bibinfo {pages} {025302} (\bibinfo {year} {2013})}\BibitemShut
  {NoStop}%
\bibitem [{\citenamefont {Brandi}, \citenamefont {Neshev},\ and\ \citenamefont
  {Ubachs}(2003)}]{brandi2003high}%
  \BibitemOpen
  \bibfield  {author} {\bibinfo {author} {\bibfnamefont {F.}~\bibnamefont
  {Brandi}}, \bibinfo {author} {\bibfnamefont {D.}~\bibnamefont {Neshev}}, \
  and\ \bibinfo {author} {\bibfnamefont {W.}~\bibnamefont {Ubachs}},\
  }\href@noop {} {\bibfield  {journal} {\bibinfo  {journal} {Phys. Rev. Lett.}\
  }\textbf {\bibinfo {volume} {91}},\ \bibinfo {pages} {163901} (\bibinfo
  {year} {2003})}\BibitemShut {NoStop}%
\bibitem [{\citenamefont {Holgado}\ \emph {et~al.}(2016)\citenamefont
  {Holgado}, \citenamefont {Hern\'andez-Garc\'ia}, \citenamefont {Alonso},
  \citenamefont {Miranda}, \citenamefont {Silva}, \citenamefont {Plaja},
  \citenamefont {Crespo},\ and\ \citenamefont {Sola}}]{holgado2016continuous}%
  \BibitemOpen
  \bibfield  {author} {\bibinfo {author} {\bibfnamefont {W.}~\bibnamefont
  {Holgado}}, \bibinfo {author} {\bibfnamefont {C.}~\bibnamefont
  {Hern\'andez-Garc\'ia}}, \bibinfo {author} {\bibfnamefont {B.}~\bibnamefont
  {Alonso}}, \bibinfo {author} {\bibfnamefont {M.}~\bibnamefont {Miranda}},
  \bibinfo {author} {\bibfnamefont {F.}~\bibnamefont {Silva}}, \bibinfo
  {author} {\bibfnamefont {L.}~\bibnamefont {Plaja}}, \bibinfo {author}
  {\bibfnamefont {H.}~\bibnamefont {Crespo}}, \ and\ \bibinfo {author}
  {\bibfnamefont {I.~J.}\ \bibnamefont {Sola}},\ }\href@noop {} {\bibfield
  {journal} {\bibinfo  {journal} {Phys. Rev. A}\ }\textbf {\bibinfo {volume}
  {93}},\ \bibinfo {pages} {013816} (\bibinfo {year} {2016})}\BibitemShut
  {NoStop}%
\bibitem [{\citenamefont {Miranda}\ \emph {et~al.}(2012)\citenamefont
  {Miranda}, \citenamefont {Arnold}, \citenamefont {Fordell}, \citenamefont
  {Silva}, \citenamefont {Alonso}, \citenamefont {Weigand}, \citenamefont
  {L’Huillier},\ and\ \citenamefont {Crespo}}]{miranda2012characterization}%
  \BibitemOpen
  \bibfield  {author} {\bibinfo {author} {\bibfnamefont {M.}~\bibnamefont
  {Miranda}}, \bibinfo {author} {\bibfnamefont {C.~L.}\ \bibnamefont {Arnold}},
  \bibinfo {author} {\bibfnamefont {T.}~\bibnamefont {Fordell}}, \bibinfo
  {author} {\bibfnamefont {F.}~\bibnamefont {Silva}}, \bibinfo {author}
  {\bibfnamefont {B.}~\bibnamefont {Alonso}}, \bibinfo {author} {\bibfnamefont
  {R.}~\bibnamefont {Weigand}}, \bibinfo {author} {\bibfnamefont
  {A.}~\bibnamefont {L’Huillier}}, \ and\ \bibinfo {author} {\bibfnamefont
  {H.}~\bibnamefont {Crespo}},\ }\href@noop {} {\bibfield  {journal} {\bibinfo
  {journal} {Opt. Express}\ }\textbf {\bibinfo {volume} {20}},\ \bibinfo
  {pages} {18732} (\bibinfo {year} {2012})}\BibitemShut {NoStop}%
\bibitem [{\citenamefont {Miranda}\ \emph {et~al.}(2017)\citenamefont
  {Miranda}, \citenamefont {Penedones}, \citenamefont {Guo}, \citenamefont
  {Harth}, \citenamefont {Louisy}, \citenamefont {Neori\v{c}i\'{c}},
  \citenamefont {L'Huillier},\ and\ \citenamefont {Arnold}}]{Miranda:17}%
  \BibitemOpen
  \bibfield  {author} {\bibinfo {author} {\bibfnamefont {M.}~\bibnamefont
  {Miranda}}, \bibinfo {author} {\bibfnamefont {J.}~\bibnamefont {Penedones}},
  \bibinfo {author} {\bibfnamefont {C.}~\bibnamefont {Guo}}, \bibinfo {author}
  {\bibfnamefont {A.}~\bibnamefont {Harth}}, \bibinfo {author} {\bibfnamefont
  {M.}~\bibnamefont {Louisy}}, \bibinfo {author} {\bibfnamefont
  {L.}~\bibnamefont {Neori\v{c}i\'{c}}}, \bibinfo {author} {\bibfnamefont
  {A.}~\bibnamefont {L'Huillier}}, \ and\ \bibinfo {author} {\bibfnamefont
  {C.~L.}\ \bibnamefont {Arnold}},\ }\href@noop {} {\bibfield  {journal}
  {\bibinfo  {journal} {J. Opt. Soc. Am. B}\ }\textbf {\bibinfo {volume}
  {34}},\ \bibinfo {pages} {190} (\bibinfo {year} {2017})}\BibitemShut
  {NoStop}%
\bibitem [{\citenamefont {P{\'e}rez-Hern{\'a}ndez}, \citenamefont {Roso},\ and\
  \citenamefont {Plaja}(2009)}]{perez2009s}%
  \BibitemOpen
  \bibfield  {author} {\bibinfo {author} {\bibfnamefont {J.}~\bibnamefont
  {P{\'e}rez-Hern{\'a}ndez}}, \bibinfo {author} {\bibfnamefont
  {L.}~\bibnamefont {Roso}}, \ and\ \bibinfo {author} {\bibfnamefont
  {L.}~\bibnamefont {Plaja}},\ }\href@noop {} {\bibfield  {journal} {\bibinfo
  {journal} {Laser Phys.}\ }\textbf {\bibinfo {volume} {19}},\ \bibinfo {pages}
  {1581} (\bibinfo {year} {2009})}\BibitemShut {NoStop}%
\bibitem [{\citenamefont {Hern\'andez-Garc\'ia}\ \emph
  {et~al.}(2010)\citenamefont {Hern\'andez-Garc\'ia}, \citenamefont
  {P\'erez-Hern\'andez}, \citenamefont {Ramos}, \citenamefont {Jarque},
  \citenamefont {Roso},\ and\ \citenamefont {Plaja}}]{hernandez2010high}%
  \BibitemOpen
  \bibfield  {author} {\bibinfo {author} {\bibfnamefont {C.}~\bibnamefont
  {Hern\'andez-Garc\'ia}}, \bibinfo {author} {\bibfnamefont {J.~A.}\
  \bibnamefont {P\'erez-Hern\'andez}}, \bibinfo {author} {\bibfnamefont
  {J.}~\bibnamefont {Ramos}}, \bibinfo {author} {\bibfnamefont {E.~C.}\
  \bibnamefont {Jarque}}, \bibinfo {author} {\bibfnamefont {L.}~\bibnamefont
  {Roso}}, \ and\ \bibinfo {author} {\bibfnamefont {L.}~\bibnamefont {Plaja}},\
  }\href@noop {} {\bibfield  {journal} {\bibinfo  {journal} {Phys. Rev. A}\
  }\textbf {\bibinfo {volume} {82}},\ \bibinfo {pages} {033432} (\bibinfo
  {year} {2010})}\BibitemShut {NoStop}%
\bibitem [{\citenamefont {Ammosov}, \citenamefont {Delone},\ and\ \citenamefont
  {Krainov}(1986)}]{adk1986}%
  \BibitemOpen
  \bibfield  {author} {\bibinfo {author} {\bibfnamefont {M.}~\bibnamefont
  {Ammosov}}, \bibinfo {author} {\bibfnamefont {N.}~\bibnamefont {Delone}}, \
  and\ \bibinfo {author} {\bibfnamefont {V.}~\bibnamefont {Krainov}},\
  }\href@noop {} {\bibfield  {journal} {\bibinfo  {journal} {Sov. Phys. JETP}\
  }\textbf {\bibinfo {volume} {64}},\ \bibinfo {pages} {1191} (\bibinfo {year}
  {1986})}\BibitemShut {NoStop}%
\bibitem [{\citenamefont {Hern{\'a}ndez-Garc{\'\i}a}\ \emph
  {et~al.}(2016)\citenamefont {Hern{\'a}ndez-Garc{\'\i}a}, \citenamefont
  {Popmintchev}, \citenamefont {Murnane}, \citenamefont {Kapteyn},
  \citenamefont {Plaja}, \citenamefont {Becker},\ and\ \citenamefont
  {Jaron-Becker}}]{hernandez2016group}%
  \BibitemOpen
  \bibfield  {author} {\bibinfo {author} {\bibfnamefont {C.}~\bibnamefont
  {Hern{\'a}ndez-Garc{\'\i}a}}, \bibinfo {author} {\bibfnamefont
  {T.}~\bibnamefont {Popmintchev}}, \bibinfo {author} {\bibfnamefont
  {M.}~\bibnamefont {Murnane}}, \bibinfo {author} {\bibfnamefont
  {H.}~\bibnamefont {Kapteyn}}, \bibinfo {author} {\bibfnamefont
  {L.}~\bibnamefont {Plaja}}, \bibinfo {author} {\bibfnamefont
  {A.}~\bibnamefont {Becker}}, \ and\ \bibinfo {author} {\bibfnamefont
  {A.}~\bibnamefont {Jaron-Becker}},\ }\href@noop {} {\bibfield  {journal}
  {\bibinfo  {journal} {New J. of Phys.}\ }\textbf {\bibinfo {volume} {18}},\
  \bibinfo {pages} {073031} (\bibinfo {year} {2016})}\BibitemShut {NoStop}%
\bibitem [{\citenamefont {Hern{\'a}ndez-Garc{\'\i}a}\ \emph
  {et~al.}(2015)\citenamefont {Hern{\'a}ndez-Garc{\'\i}a}, \citenamefont
  {Holgado}, \citenamefont {Plaja}, \citenamefont {Alonso}, \citenamefont
  {Silva}, \citenamefont {Miranda}, \citenamefont {Crespo},\ and\ \citenamefont
  {Sola}}]{hernandez2015carrier}%
  \BibitemOpen
  \bibfield  {author} {\bibinfo {author} {\bibfnamefont {C.}~\bibnamefont
  {Hern{\'a}ndez-Garc{\'\i}a}}, \bibinfo {author} {\bibfnamefont
  {W.}~\bibnamefont {Holgado}}, \bibinfo {author} {\bibfnamefont
  {L.}~\bibnamefont {Plaja}}, \bibinfo {author} {\bibfnamefont
  {B.}~\bibnamefont {Alonso}}, \bibinfo {author} {\bibfnamefont
  {F.}~\bibnamefont {Silva}}, \bibinfo {author} {\bibfnamefont
  {M.}~\bibnamefont {Miranda}}, \bibinfo {author} {\bibfnamefont
  {H.}~\bibnamefont {Crespo}}, \ and\ \bibinfo {author} {\bibfnamefont
  {I.}~\bibnamefont {Sola}},\ }\href@noop {} {\bibfield  {journal} {\bibinfo
  {journal} {Opt. Express}\ }\textbf {\bibinfo {volume} {23}},\ \bibinfo
  {pages} {21497} (\bibinfo {year} {2015})}\BibitemShut {NoStop}%
\bibitem [{\citenamefont {Hern\'andez-Garc\'ia}, \citenamefont {Sola},\ and\
  \citenamefont {Plaja}(2013)}]{hernandez2013signature}%
  \BibitemOpen
  \bibfield  {author} {\bibinfo {author} {\bibfnamefont {C.}~\bibnamefont
  {Hern\'andez-Garc\'ia}}, \bibinfo {author} {\bibfnamefont {I.~J.}\
  \bibnamefont {Sola}}, \ and\ \bibinfo {author} {\bibfnamefont
  {L.}~\bibnamefont {Plaja}},\ }\href@noop {} {\bibfield  {journal} {\bibinfo
  {journal} {Phys. Rev. A}\ }\textbf {\bibinfo {volume} {88}},\ \bibinfo
  {pages} {043848} (\bibinfo {year} {2013})}\BibitemShut {NoStop}%
\bibitem [{\citenamefont {Kretschmar}\ \emph {et~al.}(2013)\citenamefont
  {Kretschmar}, \citenamefont {Hern\'andez-Garc\'ia}, \citenamefont
  {Steingrube}, \citenamefont {Plaja}, \citenamefont {Morgner},\ and\
  \citenamefont {Kova\v{c}ev}}]{kretschmar2013spatial}%
  \BibitemOpen
  \bibfield  {author} {\bibinfo {author} {\bibfnamefont {M.}~\bibnamefont
  {Kretschmar}}, \bibinfo {author} {\bibfnamefont {C.}~\bibnamefont
  {Hern\'andez-Garc\'ia}}, \bibinfo {author} {\bibfnamefont {D.~S.}\
  \bibnamefont {Steingrube}}, \bibinfo {author} {\bibfnamefont
  {L.}~\bibnamefont {Plaja}}, \bibinfo {author} {\bibfnamefont
  {U.}~\bibnamefont {Morgner}}, \ and\ \bibinfo {author} {\bibfnamefont
  {M.}~\bibnamefont {Kova\v{c}ev}},\ }\href@noop {} {\bibfield  {journal}
  {\bibinfo  {journal} {Phys. Rev. A}\ }\textbf {\bibinfo {volume} {88}},\
  \bibinfo {pages} {013805} (\bibinfo {year} {2013})}\BibitemShut {NoStop}%
\bibitem [{\citenamefont {Hickstein}\ \emph {et~al.}(2015)\citenamefont
  {Hickstein}, \citenamefont {Dollar}, \citenamefont {Grychtol}, \citenamefont
  {Ellis}, \citenamefont {Knut}, \citenamefont {Hernández-García},
  \citenamefont {Zusin}, \citenamefont {Gentry}, \citenamefont {Shaw},
  \citenamefont {Fan}, \citenamefont {Dorney}, \citenamefont {Becker},
  \citenamefont {andHenry C.~Kapteyn}, \citenamefont {Murnane},\ and\
  \citenamefont {Durfee}}]{hickstein2015non}%
  \BibitemOpen
  \bibfield  {author} {\bibinfo {author} {\bibfnamefont {D.~D.}\ \bibnamefont
  {Hickstein}}, \bibinfo {author} {\bibfnamefont {F.~J.}\ \bibnamefont
  {Dollar}}, \bibinfo {author} {\bibfnamefont {P.}~\bibnamefont {Grychtol}},
  \bibinfo {author} {\bibfnamefont {J.~L.}\ \bibnamefont {Ellis}}, \bibinfo
  {author} {\bibfnamefont {R.}~\bibnamefont {Knut}}, \bibinfo {author}
  {\bibfnamefont {C.}~\bibnamefont {Hernández-García}}, \bibinfo {author}
  {\bibfnamefont {D.}~\bibnamefont {Zusin}}, \bibinfo {author} {\bibfnamefont
  {C.}~\bibnamefont {Gentry}}, \bibinfo {author} {\bibfnamefont {J.~M.}\
  \bibnamefont {Shaw}}, \bibinfo {author} {\bibfnamefont {T.}~\bibnamefont
  {Fan}}, \bibinfo {author} {\bibfnamefont {K.~M.}\ \bibnamefont {Dorney}},
  \bibinfo {author} {\bibfnamefont {A.}~\bibnamefont {Becker}}, \bibinfo
  {author} {\bibfnamefont {A.~J.-B.}\ \bibnamefont {andHenry C.~Kapteyn}},
  \bibinfo {author} {\bibfnamefont {M.~M.}\ \bibnamefont {Murnane}}, \ and\
  \bibinfo {author} {\bibfnamefont {C.~G.}\ \bibnamefont {Durfee}},\
  }\href@noop {} {\bibfield  {journal} {\bibinfo  {journal} {Nat. Photon.}\ }
  (\bibinfo {year} {2015})}\BibitemShut {NoStop}%
\bibitem [{\citenamefont {Hern\'andez-Garc\'ia}\ \emph
  {et~al.}(2013)\citenamefont {Hern\'andez-Garc\'ia}, \citenamefont {Pic\'on},
  \citenamefont {San~Rom\'an},\ and\ \citenamefont
  {Plaja}}]{hernandez2013attosecond}%
  \BibitemOpen
  \bibfield  {author} {\bibinfo {author} {\bibfnamefont {C.}~\bibnamefont
  {Hern\'andez-Garc\'ia}}, \bibinfo {author} {\bibfnamefont {A.}~\bibnamefont
  {Pic\'on}}, \bibinfo {author} {\bibfnamefont {J.}~\bibnamefont
  {San~Rom\'an}}, \ and\ \bibinfo {author} {\bibfnamefont {L.}~\bibnamefont
  {Plaja}},\ }\href@noop {} {\bibfield  {journal} {\bibinfo  {journal} {Phys.
  Rev. Lett.}\ }\textbf {\bibinfo {volume} {111}},\ \bibinfo {pages} {083602}
  (\bibinfo {year} {2013})}\BibitemShut {NoStop}%
\bibitem [{\citenamefont {Conejero~Jarque}\ \emph {et~al.}(2017)\citenamefont
  {Conejero~Jarque}, \citenamefont {San~Rom\'an}, \citenamefont {Silva},
  \citenamefont {Romero}, \citenamefont {Holgado}, \citenamefont
  {Gonz\'alez-Galicia}, \citenamefont {Sola},\ and\ \citenamefont
  {Cres\-po}}]{conerejo2016tod}%
  \BibitemOpen
  \bibfield  {author} {\bibinfo {author} {\bibfnamefont {E.}~\bibnamefont
  {Conejero~Jarque}}, \bibinfo {author} {\bibfnamefont {J.}~\bibnamefont
  {San~Rom\'an}}, \bibinfo {author} {\bibfnamefont {F.}~\bibnamefont {Silva}},
  \bibinfo {author} {\bibfnamefont {R.}~\bibnamefont {Romero}}, \bibinfo
  {author} {\bibfnamefont {W.}~\bibnamefont {Holgado}}, \bibinfo {author}
  {\bibfnamefont {M.~A.}\ \bibnamefont {Gonz\'alez-Galicia}}, \bibinfo {author}
  {\bibfnamefont {I.}~\bibnamefont {Sola}}, \ and\ \bibinfo {author}
  {\bibfnamefont {H.}~\bibnamefont {Cres\-po}},\ }\href@noop {} {\bibfield
  {journal} {\bibinfo  {journal} {submitted}\ } (\bibinfo {year}
  {2017})}\BibitemShut {NoStop}%
\bibitem [{\citenamefont {Fabris}\ \emph {et~al.}(2015)\citenamefont {Fabris},
  \citenamefont {Holgado}, \citenamefont {Silva}, \citenamefont {Witting},
  \citenamefont {Tisch},\ and\ \citenamefont {Cres\-po}}]{fabris2015single}%
  \BibitemOpen
  \bibfield  {author} {\bibinfo {author} {\bibfnamefont {D.}~\bibnamefont
  {Fabris}}, \bibinfo {author} {\bibfnamefont {W.}~\bibnamefont {Holgado}},
  \bibinfo {author} {\bibfnamefont {F.}~\bibnamefont {Silva}}, \bibinfo
  {author} {\bibfnamefont {T.}~\bibnamefont {Witting}}, \bibinfo {author}
  {\bibfnamefont {J.~W.}\ \bibnamefont {Tisch}}, \ and\ \bibinfo {author}
  {\bibfnamefont {H.}~\bibnamefont {Cres\-po}},\ }\href@noop {} {\bibfield
  {journal} {\bibinfo  {journal} {Opt. Express}\ }\textbf {\bibinfo {volume}
  {23}},\ \bibinfo {pages} {32803} (\bibinfo {year} {2015})}\BibitemShut
  {NoStop}%
\bibitem [{\citenamefont {Heyl}\ \emph {et~al.}(2016)\citenamefont {Heyl},
  \citenamefont {Coudert-Alteirac}, \citenamefont {Miranda}, \citenamefont
  {Louisy}, \citenamefont {Kovacs}, \citenamefont {Tosa}, \citenamefont
  {Balogh}, \citenamefont {Varj{\'u}}, \citenamefont {L’Huillier},
  \citenamefont {Couairon},\ and\ \citenamefont {Arnold}}]{heyl2016scale}%
  \BibitemOpen
  \bibfield  {author} {\bibinfo {author} {\bibfnamefont {C.~M.}\ \bibnamefont
  {Heyl}}, \bibinfo {author} {\bibfnamefont {H.}~\bibnamefont
  {Coudert-Alteirac}}, \bibinfo {author} {\bibfnamefont {M.}~\bibnamefont
  {Miranda}}, \bibinfo {author} {\bibfnamefont {M.}~\bibnamefont {Louisy}},
  \bibinfo {author} {\bibfnamefont {K.}~\bibnamefont {Kovacs}}, \bibinfo
  {author} {\bibfnamefont {V.}~\bibnamefont {Tosa}}, \bibinfo {author}
  {\bibfnamefont {E.}~\bibnamefont {Balogh}}, \bibinfo {author} {\bibfnamefont
  {K.}~\bibnamefont {Varj{\'u}}}, \bibinfo {author} {\bibfnamefont
  {A.}~\bibnamefont {L’Huillier}}, \bibinfo {author} {\bibfnamefont
  {A.}~\bibnamefont {Couairon}}, \ and\ \bibinfo {author} {\bibfnamefont
  {C.~L.}\ \bibnamefont {Arnold}},\ }\href@noop {} {\bibfield  {journal}
  {\bibinfo  {journal} {Optica}\ }\textbf {\bibinfo {volume} {3}},\ \bibinfo
  {pages} {75} (\bibinfo {year} {2016})}\BibitemShut {NoStop}%
\bibitem [{\citenamefont {B{\"o}hle}\ \emph {et~al.}(2014)\citenamefont
  {B{\"o}hle}, \citenamefont {Kretschmar}, \citenamefont {Jullien},
  \citenamefont {Kovacs}, \citenamefont {Miranda}, \citenamefont {Romero},
  \citenamefont {Crespo}, \citenamefont {Morgner}, \citenamefont {Simon},
  \citenamefont {Lopez-Martens},\ and\ \citenamefont
  {T. Nagy}}]{bohle2014compression}%
  \BibitemOpen
  \bibfield  {author} {\bibinfo {author} {\bibfnamefont {F.}~\bibnamefont
  {B{\"o}hle}}, \bibinfo {author} {\bibfnamefont {M.}~\bibnamefont
  {Kretschmar}}, \bibinfo {author} {\bibfnamefont {A.}~\bibnamefont {Jullien}},
  \bibinfo {author} {\bibfnamefont {M.}~\bibnamefont {Kovacs}}, \bibinfo
  {author} {\bibfnamefont {M.}~\bibnamefont {Miranda}}, \bibinfo {author}
  {\bibfnamefont {R.}~\bibnamefont {Romero}}, \bibinfo {author} {\bibfnamefont
  {H.}~\bibnamefont {Crespo}}, \bibinfo {author} {\bibfnamefont
  {U.}~\bibnamefont {Morgner}}, \bibinfo {author} {\bibfnamefont
  {P.}~\bibnamefont {Simon}}, \bibinfo {author} {\bibfnamefont
  {R.}~\bibnamefont {Lopez-Martens}}, \ and\ \bibinfo {author} {\bibnamefont
  {T. Nagy}},\ }\href@noop {} {\bibfield  {journal} {\bibinfo  {journal}
  {Laser Phys. Lett.}\ }\textbf {\bibinfo {volume} {11}},\ \bibinfo {pages}
  {095401} (\bibinfo {year} {2014})}\BibitemShut {NoStop}%
\bibitem [{\citenamefont {Karimi}\ \emph {et~al.}(2013)\citenamefont {Karimi},
  \citenamefont {Altucci}, \citenamefont {Tosa}, \citenamefont {Velotta},\ and\
  \citenamefont {Marrucci}}]{Karimi:13}%
  \BibitemOpen
  \bibfield  {author} {\bibinfo {author} {\bibfnamefont {E.}~\bibnamefont
  {Karimi}}, \bibinfo {author} {\bibfnamefont {C.}~\bibnamefont {Altucci}},
  \bibinfo {author} {\bibfnamefont {V.}~\bibnamefont {Tosa}}, \bibinfo {author}
  {\bibfnamefont {R.}~\bibnamefont {Velotta}}, \ and\ \bibinfo {author}
  {\bibfnamefont {L.}~\bibnamefont {Marrucci}},\ }\href {\doibase
  10.1364/OE.21.024991} {\bibfield  {journal} {\bibinfo  {journal} {Opt.
  Express}\ }\textbf {\bibinfo {volume} {21}},\ \bibinfo {pages} {24991}
  (\bibinfo {year} {2013})}\BibitemShut {NoStop}%
\bibitem [{\citenamefont {Wang}\ \emph {et~al.}(2008)\citenamefont {Wang},
  \citenamefont {Wu}, \citenamefont {Li}, \citenamefont {Mashiko},
  \citenamefont {Gilbertson},\ and\ \citenamefont
  {Chang}}]{wang2008generation}%
  \BibitemOpen
  \bibfield  {author} {\bibinfo {author} {\bibfnamefont {H.}~\bibnamefont
  {Wang}}, \bibinfo {author} {\bibfnamefont {Y.}~\bibnamefont {Wu}}, \bibinfo
  {author} {\bibfnamefont {C.}~\bibnamefont {Li}}, \bibinfo {author}
  {\bibfnamefont {H.}~\bibnamefont {Mashiko}}, \bibinfo {author} {\bibfnamefont
  {S.}~\bibnamefont {Gilbertson}}, \ and\ \bibinfo {author} {\bibfnamefont
  {Z.}~\bibnamefont {Chang}},\ }\href@noop {} {\bibfield  {journal} {\bibinfo
  {journal} {Opt. Express}\ }\textbf {\bibinfo {volume} {16}},\ \bibinfo
  {pages} {14448} (\bibinfo {year} {2008})}\BibitemShut {NoStop}%
\bibitem [{\citenamefont {Rudawski}\ \emph {et~al.}(2015)\citenamefont
  {Rudawski}, \citenamefont {Harth}, \citenamefont {Guo}, \citenamefont
  {Lorek}, \citenamefont {Miranda}, \citenamefont {Heyl}, \citenamefont
  {Larsen}, \citenamefont {Ahrens}, \citenamefont {Prochnow}, \citenamefont
  {Binhammer}, \citenamefont {Morgner}, \citenamefont {Mauritsson},
  \citenamefont {L’Huillier},\ and\ \citenamefont
  {Arnold}}]{rudawski2015carrier}%
  \BibitemOpen
  \bibfield  {author} {\bibinfo {author} {\bibfnamefont {P.}~\bibnamefont
  {Rudawski}}, \bibinfo {author} {\bibfnamefont {A.}~\bibnamefont {Harth}},
  \bibinfo {author} {\bibfnamefont {C.}~\bibnamefont {Guo}}, \bibinfo {author}
  {\bibfnamefont {E.}~\bibnamefont {Lorek}}, \bibinfo {author} {\bibfnamefont
  {M.}~\bibnamefont {Miranda}}, \bibinfo {author} {\bibfnamefont {C.~M.}\
  \bibnamefont {Heyl}}, \bibinfo {author} {\bibfnamefont {E.~W.}\ \bibnamefont
  {Larsen}}, \bibinfo {author} {\bibfnamefont {J.}~\bibnamefont {Ahrens}},
  \bibinfo {author} {\bibfnamefont {O.}~\bibnamefont {Prochnow}}, \bibinfo
  {author} {\bibfnamefont {T.}~\bibnamefont {Binhammer}}, \bibinfo {author}
  {\bibfnamefont {U.}~\bibnamefont {Morgner}}, \bibinfo {author} {\bibfnamefont
  {J.}~\bibnamefont {Mauritsson}}, \bibinfo {author} {\bibfnamefont
  {A.}~\bibnamefont {L’Huillier}}, \ and\ \bibinfo {author} {\bibfnamefont
  {C.~L.}\ \bibnamefont {Arnold}},\ }\href@noop {} {\bibfield  {journal}
  {\bibinfo  {journal} {The European Phys. J. D}\ }\textbf {\bibinfo {volume}
  {69}},\ \bibinfo {pages} {1} (\bibinfo {year} {2015})}\BibitemShut {NoStop}%
\bibitem [{\citenamefont {Timmers}\ \emph {et~al.}(2017)\citenamefont
  {Timmers}, \citenamefont {Kobayashi}, \citenamefont {Chang}, \citenamefont
  {Reduzzi}, \citenamefont {Neumark},\ and\ \citenamefont
  {Leone}}]{timmers2017postcompression}%
  \BibitemOpen
  \bibfield  {author} {\bibinfo {author} {\bibfnamefont {H.}~\bibnamefont
  {Timmers}}, \bibinfo {author} {\bibfnamefont {Y.}~\bibnamefont {Kobayashi}},
  \bibinfo {author} {\bibfnamefont {K.~F.}\ \bibnamefont {Chang}}, \bibinfo
  {author} {\bibfnamefont {M.}~\bibnamefont {Reduzzi}}, \bibinfo {author}
  {\bibfnamefont {D.~M.}\ \bibnamefont {Neumark}}, \ and\ \bibinfo {author}
  {\bibfnamefont {S.~R.}\ \bibnamefont {Leone}},\ }\href@noop {} {\bibfield
  {journal} {\bibinfo  {journal} {Optics Letters}\ }\textbf {\bibinfo {volume}
  {42}},\ \bibinfo {pages} {811} (\bibinfo {year} {2017})}\BibitemShut
  {NoStop}%
\end{thebibliography}%
	\bibliographystyle{aipnum4-1.bst}

\end{document}